\documentclass[10pt,english]{article}
\usepackage{lmodern}

\usepackage[T1]{fontenc}
\usepackage[latin9]{inputenc}
\usepackage[letterpaper]{geometry}
\usepackage{babel}
\usepackage{array}
\usepackage{float}
\usepackage{multirow}
\usepackage{amsmath}
\usepackage{amssymb}
\usepackage{graphicx}
\usepackage{esint}
\usepackage{subscript}
\usepackage{float}
\usepackage{color}
\usepackage[bookmarks=false,linkcolor=red]{hyperref}
\hypersetup{colorlinks=true,breaklinks,urlcolor=[rgb]{0,0,1},citecolor=[rgb]{0,0.6,0}}
\usepackage{url}
\usepackage{multirow} 
\usepackage{array}
\usepackage[font=footnotesize,format=plain,labelfont=bf, margin=0.8cm]{caption}
\usepackage{url}
\usepackage{multirow}
\usepackage{array}
\usepackage{booktabs}
\usepackage{rotating}
\usepackage[numbers]{natbib}
\usepackage{wrapfig}
\makeatletter

\providecommand{\tabularnewline}{\\}
\floatstyle{ruled}
\newfloat{algorithm}{tbp}{loa}
\providecommand{\algorithmname}{Algorithm}
\floatname{algorithm}{\protect\algorithmname}

\@ifundefined{date}{}{\date{}}
\date{}

\makeatother

\begin{document}

\title{Shrunken Locally Linear Embedding for Passive Microwave Retrieval
of Precipitation}
\author{{\normalsize{Ardeshir M. Ebtehaj}}%
\thanks{School of Civil and Environmental Engineering, Georgia Institute of
Technology, Atlanta GA (e-mail: \href{mailto:mebtehaj@gatech.edu}{mebtehaj@gatech.edu}; \href{mailto:rlbras@gatech.edu}{rlbras@gatech.edu})%
}{\normalsize{ $\qquad$Rafael L. Bras$^{*}$$\qquad$Efi Foufoula-Georgiou}}%
\thanks{Department of Civil Engineering, University of Minnesota, Minneapolis
MN (e-mail: \href{mailto:efi@umn.edu}{efi@umn.edu})%
}}
\maketitle

\begin{abstract}
\textbf{This paper introduces a new Bayesian approach to the inverse problem
of passive microwave rainfall retrieval. The proposed methodology
relies on a regularization technique and makes use of two joint dictionaries
of coincidental rainfall profiles and their corresponding upwelling
spectral radiative fluxes. A sequential detection-estimation strategy
is adopted, which basically assumes that similar rainfall intensity
values and their spectral radiances live close to some sufficiently
smooth manifolds with analogous local geometry. The detection step
employs a nearest neighborhood classification rule, while the estimation
scheme is equipped with a constrained shrinkage estimator to ensure
stability of retrieval and some physical consistency. The algorithm
is examined using coincidental observations of the active precipitation
radar (PR) and passive microwave imager (TMI) on board the Tropical
Rainfall Measuring Mission (TRMM) satellite. We present promising
results of instantaneous rainfall retrieval for some tropical storms
and mesoscale convective systems over ocean, land, and coastal zones.
We provide evidence that the algorithm is capable of properly capturing
different storm morphologies including high intensity rain-cells and
trailing light rainfall, especially over land and coastal areas. The
algorithm is also validated at an annual scale for calendar year 2013
versus the standard (version 7) radar (2A25) and radiometer (2A12)
rainfall products of the TRMM satellite.}
\end{abstract}

\section{INTRODUCTION \label{sec:1}}

From a mathematical standpoint, rainfall retrieval from remotely sensed
observations is an inverse problem in which we aim to estimate the
rainfall intensity from its indirect and noisy measurements. Passive
retrieval of rainfall from upwelling spectral radiances is one of
the most challenging atmospheric retrieval problems, chiefly because
the rainfall spectral signatures are often downsampled, significantly
corrupted with the background radiation and are non-linearly related
to the rainfall vertical profile. Retrieval of rainfall from visible
and infrared observations typically relies on empirical approaches
as the measurements only respond to the radiative fluxes from the
upper portion of the cloud layers \citep[, among others]{Bar70,Bla75,KilR76,Ark79,ArkA89,LovA79,HsuGSG97}.
In the microwave wavelengths ($\sim$ 6-to-200 GHz), the hydrometeor
vertical profile is optically active and alters the upwelling radiations
in the entire atmospheric column through absorption-emission and scattering
processes. Over ocean, absorption-emission of the atmospheric liquid
water can be well distinguished from the cold background by the physical
laws of radiative transfer \citep[e.g.,][]{Wil77,Liu1992}. In addition,
the attenuation of the polarized ocean surface emission by atmospheric
hydrometeors \citep[e.g.,][and references therein]{Petty1990,petty94a,petty94b,PetL13a}
and scattering by ice particles \citep[e.g.,][]{Spencer1986,Liu1992}
also give rise to high signal-to-noise ratio in the rainfall spectral
signatures making the retrieval problem more straightforward over
ocean than over land. Over land, radiation from the highly emissive
heterogeneous land surfaces often masks the hydrometeor emission signal
enforcing the retrieval approaches to rely mostly on the complex scattering
effects of the ice particles in the raining clouds \citep{Wilheit1994,PETTY1996,WilKF03}.
As a result of these major differences over ocean and land, two classes
of physically-based and empirical microwave retrieval algorithms have
emerged. The empirical approaches have been predominantly used for
retrieval over land while physically-based methods been used over
ocean.

Over ocean, physically-based methods typically follow two distinct
strategies. The first family of these algorithms \citep[e.g.,][]{Wil77}
simplifies the basic radiative transfer equation for atmospheric constituents
under the axially symmetric scattering and Rayleigh-Jeans approximation.
Given the observed spectral radiative fluxes with minimal scattering
effect, the simplified equations permit to obtain atmospheric absorptivity,
dropsize distribution and thus the rainfall intensity profile. The
second class of methodologies \citep[among others]{Olson89,Mugnai1993,KumG94a,raey,Evan95,KumOWG96,Kumetal01,KumRCRB10},
known as the Bayesian retrieval approaches, exploit a statistically
representative a priori generated database that encodes the correspondence
between the spectral brightness temperatures and rainfall profiles.
In physically generated databases the causal relationships between
the precipitation profiles and their upwelling spectral radiances
are modeled using a combination of cloud resolving and radiative transfer
models. Sophisticated numerical cloud resolving models (e.g., Goddard
Cumulus Ensemble Model) are being used to produce a large collection
of raining and non-raining cloud structures with distinct hydrometeor
profiles. Then, for all of these profiles, a radiative transfer model
is employed to obtain their spectral radiances at the top of the atmosphere.
Finally, this database is utilized to retrieve rainfall profiles from
observed microwave radiances using an inversion scheme. This approach
has been the corner stone of the Goddard Profiling Algorithm GPROF
\citep{KumOWG96,Kumetal01,KumRCRB10} used to produce the TRMM operational
passive retrieval products. On the other hand, over land, empirical
methods typically rely on a scattering index \citep{Spenc83,Gro91},
which relates the depression in the high-frequency channels (e.g.,
85 GHz) to the surface rainfall, in response to the frozen hydrometeors
commonly found in the raining clouds. The magnitude of the high-frequency
depression is naturally not independent of the land surface emissivity.
As a result, prior to the rainfall estimation, different screening
approaches are commonly employed to properly exclude depressions caused
by the background noise (e.g., snow and desert surfaces). Among these,
the early version of the GPROF \citep{KumOWG96,Kumetal01} suggests
a static thresholding (22-85 GHz > 8 Kelvin) to detect raining signatures
of the spectral brightness temperatures measured by the TRMM microwave
imager (TMI). A more involved scattering index has also been suggested
by \citet{Ferraro1994,Ferraro1998}, which has been partly used to
develop the launch version of the land retrieval algorithm for the
Advanced Microwave Scanning Radiometer---Earth Observing System (AMSR-E)
\citep{WilKF03}.

Upon successful launch of the TRMM satellite, a major body of research
has also been devoted to developing rainfall retrieval algorithms
by exploiting the coincidental observations provided by the TMI and
TRMM precipitation radar (PR) \citep[e.g.,][]{Olson1996,Had97,MarZ98,GreA02,Greetal04,GreO06,Skofronick-Jackson2003}.
The basic idea has been focused on combining, in an optimal sense,
the information content of both sensors for obtaining improved estimates
of the rainfall profile and perhaps micro-physical properties of the
atmospheric constituents. Typically, these methods use a variational
cost function to reconcile the observations provided by both instruments
\citep[e.g.,][]{GreA02,Greetal04,GreO06,Skofronick-Jackson2003},
while recently \citet{KumRCRB10} combined the PR data with the physically-driven
database of the GPROF algorithm to make the database more observationally
consistent. Using coincidental TMI and PR observations \citet{PetL13a,PetL13b}
introduced a low-dimensional approximation method, using Principal
Component Analysis (PCA), known as the University of Wisconsin (UW)
algorithm. Specifically, \citet{PetL13a} suggested a PCA based approach
to project the nine TMI channels onto three pseudo channels for filtering
the background noise and reducing redundancies in the TMI channels.
These pseudo channels are then used within a matching process to efficiently
recoPhysical retrievals of over-ocean rain rate from multichannel
microwavever the surface rainfall from a compactly designed a priori
database in a Bayesian context.

Passive rainfall retrieval remains a challenge especially for: 1)
detection and estimation of the light rainfall over land and adjacent
to coastlines, 2) unbiased estimation of rainfall over highly emissive
and nonhomogeneous land surfaces, 3) probabilistic recovery of small-scale
features of the rainfall extremes both over land and ocean \citep[see, e.g.,][ and references therein]{McCollum2005,PetL13a}.
In this paper, motivated by these continuous challenges, we introduce
a new Bayesian retrieval algorithm, called Shrunken Locally Linear
Embedding for Retrieval of Precipitation (ShARP). This retrieval algorithm
is guided by a priori collections of spectral radiances and their
corresponding rainfall profiles, so-called spectral and rainfall ``dictionaries''.
The core part is inspired by the concept of locally linear embedding
\citep{RowSL00}, which assumes that ``similar'' spectral radiances
and their corresponding rainfall profiles live close to two joint
smooth manifolds allowing locally linear approximations. To retrieve
rainfall, ShARP uses a $k$-nearest neighborhood classification (detection
step) coupled with a modern shrinkage regularization scheme (estimation
step). For an observed spectral radiance, the detection step finds
similar signatures in the spectral dictionary and decides whether
the observed spectral radiance is non-raining or raining. For a raining
spectral radiance, the estimation step uses a shrinkage estimator
to obtain its representation coefficients in the spectral dictionary.
Then, the representation coefficients are used to combine the corresponding
rainfall profiles from the rainfall dictionary to result in the rainfall
retrieval. 

In summary, the main contribution and advantageous features of this
algorithm for addressing the aforementioned retrieval challenges are:
1) The use of supervised nearest neighborhood classification results
in minimal sensitivity to the variability of the underlying land surface
emissivity. This property promises improved retrieval over troublesome
surfaces and coastal zones. 2) The core estimation step makes use
of a modern constrained regularization scheme giving rise to sufficiently
stable retrievals with reduced error, compared to the classic least-squares
solutions. 3) By design and due to the used regularization scheme,
the algorithm is flexible and robust enough to employ dictionaries
populated either empirically or via physically-based modeling or a
combination of them. 4) The algorithm allows us to approximate the
posterior probability density function of the retrieved rainfall,
especially useful for hazard assessment of the rainfall extremes and
their hydro-geomorphic impacts. It is important to note that the current
implementation of our algorithm is fully empirical, as we populate
the rainfall and spectral dictionaries only with the coincidental
observations of the TRMM-PR and TMI. Therefore, in the absence of
any independent ground-based validation, all of the presented retrieval
results are bounded by accuracy of the PR sensor/algorithm \citep[see,][]{Berg2006}.
Clearly, as we validate our results with the 2A25, improved retrievals
often do not come as a surprise; however, they remain of significant
importance as currently the passive retrieval methods are empirical
over land and coastal areas. 

Section \ref{sec: 2} is devoted to explaining the rainfall dataset
and studying the rainfall spectral patterns relevant to the design
of the presented algorithm. Section \ref{sec:3} explains the details
of the ShARP algorithm. Using the TRMM data, in Section \ref{sec:4},
some retrieval results are presented and compared with the currently
operational PR-2A25 and TMI-2A12 retrieval products (version 7). Conclusions
are drawn and future lines of research are pointed out in Section
\ref{sec:5}.

\section{TRMM Rainfall Database \label{sec: 2} }

Before we embark upon a detailed algorithmic discussion, we provide
a brief explanation of the dataset used and some relevant insights
into the structure of raining and non-raining microwave spectral patterns,
which are essential to the development of our algorithm. 

The TRMM-PR is a Ku band radar that operates in a single polarization
mode at frequency 13.8 GHz. Currently, the PR provides direct measurements
of rainfall reflectivity at grid size 4-to-5 km over a swath width
of 247 km and samples the first 15 to 20 km of the troposphere at
every \textasciitilde{}250 m at nadir. On the other hand, TMI is a
dual-polarized multichannel radiometer that operates on central frequencies
10.65, 19.35, 21.3, 37.0, and 85.5 GHz. All of the channels are horizontally
and vertically polarized except the vertical water vapor channel 21.3
GHz. Currently, TMI provides spectral brightness temperatures over
a swath width of 878 km with nominal spatial resolution greater than
5.1 km at 85.5 GHz. By design, the TMI and PR sensors provide overlapping
observations over the inner swath within the radar field of view at
different resolutions. A thorough exposition of the TRMM sensor packages
can be found in \citep{Kum98}. 
\begin{figure}[h]
\noindent \begin{centering}
\includegraphics[width=3.15in]{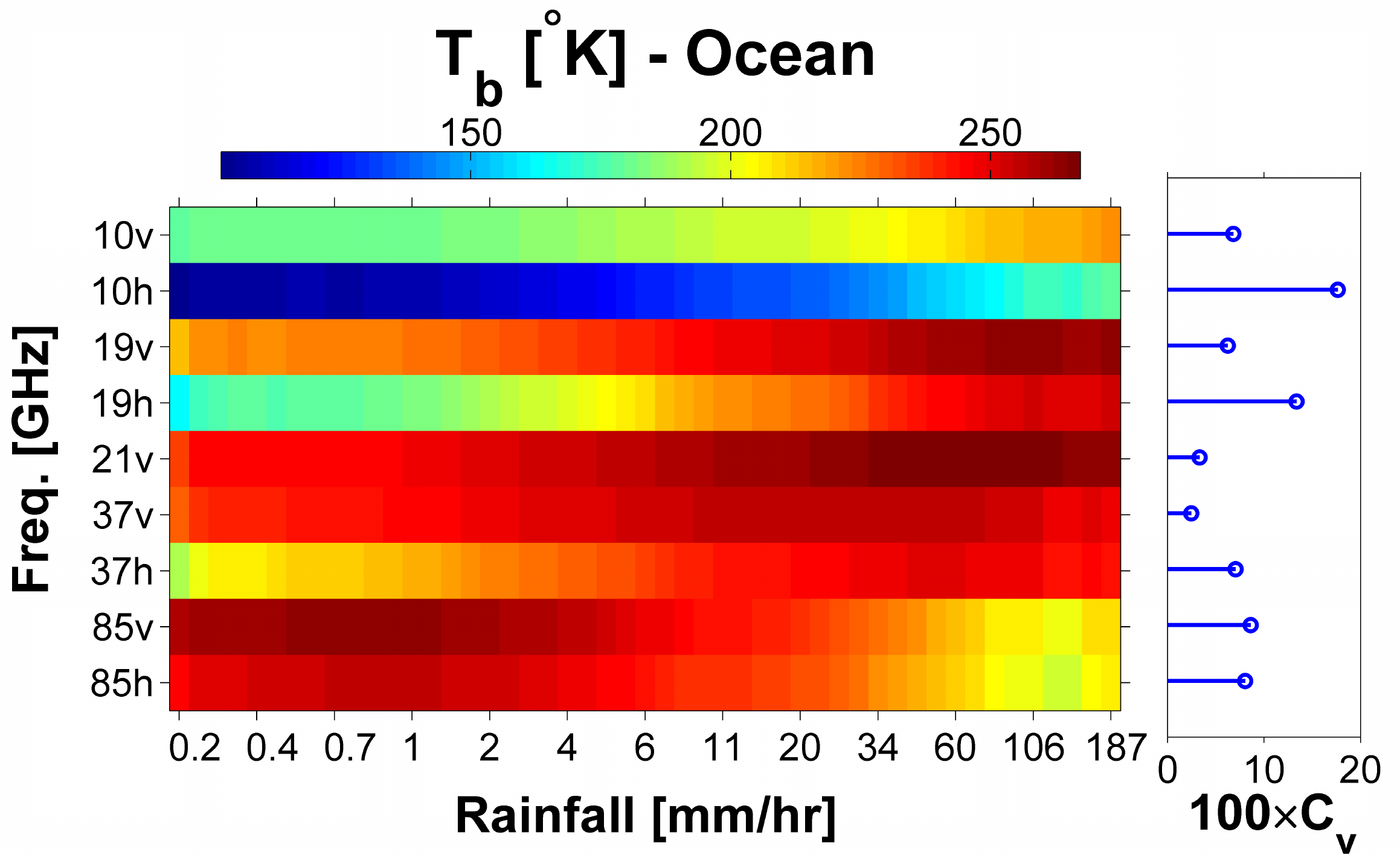}\hspace{0.1in}\includegraphics[width=3.15in]{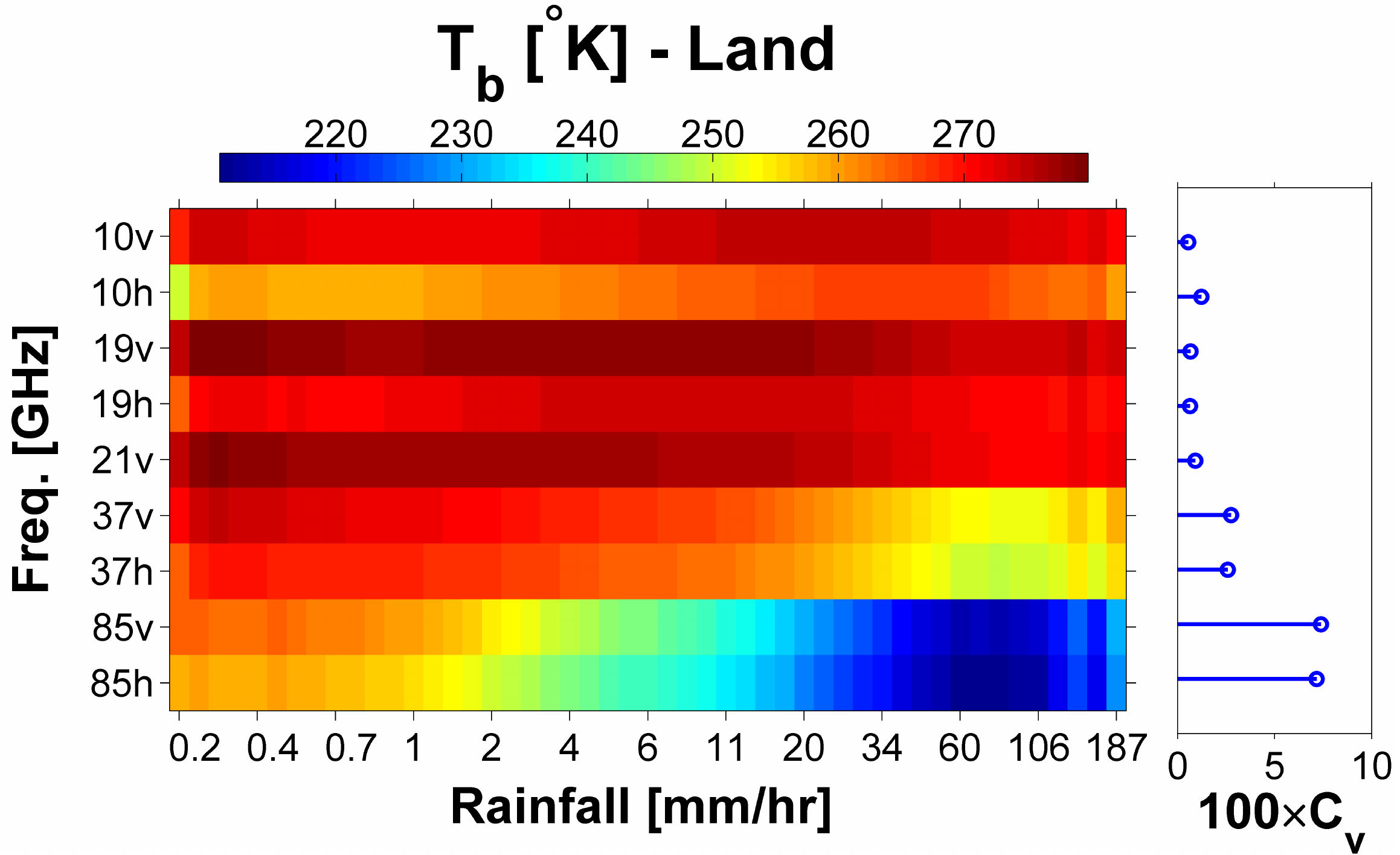} 
\par\end{centering}

\protect\caption{Expected values of the spectral brightness temperatures for different
intervals of the surface rainfall intensity over ocean (left panel)
and land (right panel). The images are inferred from coincidental
pairs of the TMI-1B11 and PR-2A25 products obtained from 1000 randomly
chosen orbits in our rainfall database. The stem plots demonstrate
the coefficients of variation for each spectral band in response to
the underlying rainfall variability. Note that, the rainfall intervals
on the x-axis are logarithmically spaced between 0.2 to 200 mm/hr.
\label{fig:1} }
\end{figure}

Here, we use the coincidental 2A25 (level-II) product of the radar
profiling algorithm \citep{IguKMA00} and the 1B11 (level-I) product
of the radiometer to construct the rainfall and spectral dictionaries.
To register all of the data onto a single grid of latitude/longitude,
we simply used the nearest neighborhood interpolation and mapped the
TMI spectral temperatures onto the reported PR grids. Note that in
this case, we neither loose nor add any information and retrieve rainfall
at the native resolution of the 1B11 at the high-frequency channel
85 GHz. Clearly, in this resolution, the lower frequency channels
provide redundant spectral information over neighboring grid-boxes
while their combinations with higher frequency channels may still
provide distinct multi-spectral information. Accordingly, throughout
this paper, we use a large collection of collocated TMI and PR data,
hereafter called ``rainfall database'', over the TRMM inner swath
for all orbital tracks in calendar years 2002, 2005, 2008, 2011, and
2012. 

Using the collected rainfall database, Fig.~\ref{fig:1} shows the
conditional expectations of the TMI spectral brightness temperatures
for different ranges of the PR rainfall intensities as well as their
coefficients of variation. Specifically, each column of the shown
images demonstrates the conditional mean of the TMI channels, while
each row shows the average response of the channels to the underlying
rainfall variability. On the other hand, the stem plots represent
the coefficients of variation of the brightness temperatures for each
channel. Over ocean (left panel), we see that almost all frequencies
are relatively responsive to the underlying surface rainfall variability.
Horizontal channels of 10 and 19 GHz show the maximum normalized variations,
while the vertical polarizations in frequencies of 21 and 37 GHz are
the least responsive channels. This observation is consistent with
the fact that ocean surface is less emissive in horizontal polarizations
for the TMI view angle, giving rise to colder background and thus
larger signal-to-noise ratio of the raining signatures \citep{Wil79}.
On the contrary, over land, almost all of the low-frequency channels
below 21 GHz show relatively small coefficients of variation compared
to the higher frequencies. It will be clear later on that these coefficients
of variation can be used to properly weight each channel to better
guide the proposed retrieval approach. 

\begin{figure}[t]
\noindent \begin{centering}
\includegraphics[width=0.77\paperwidth]{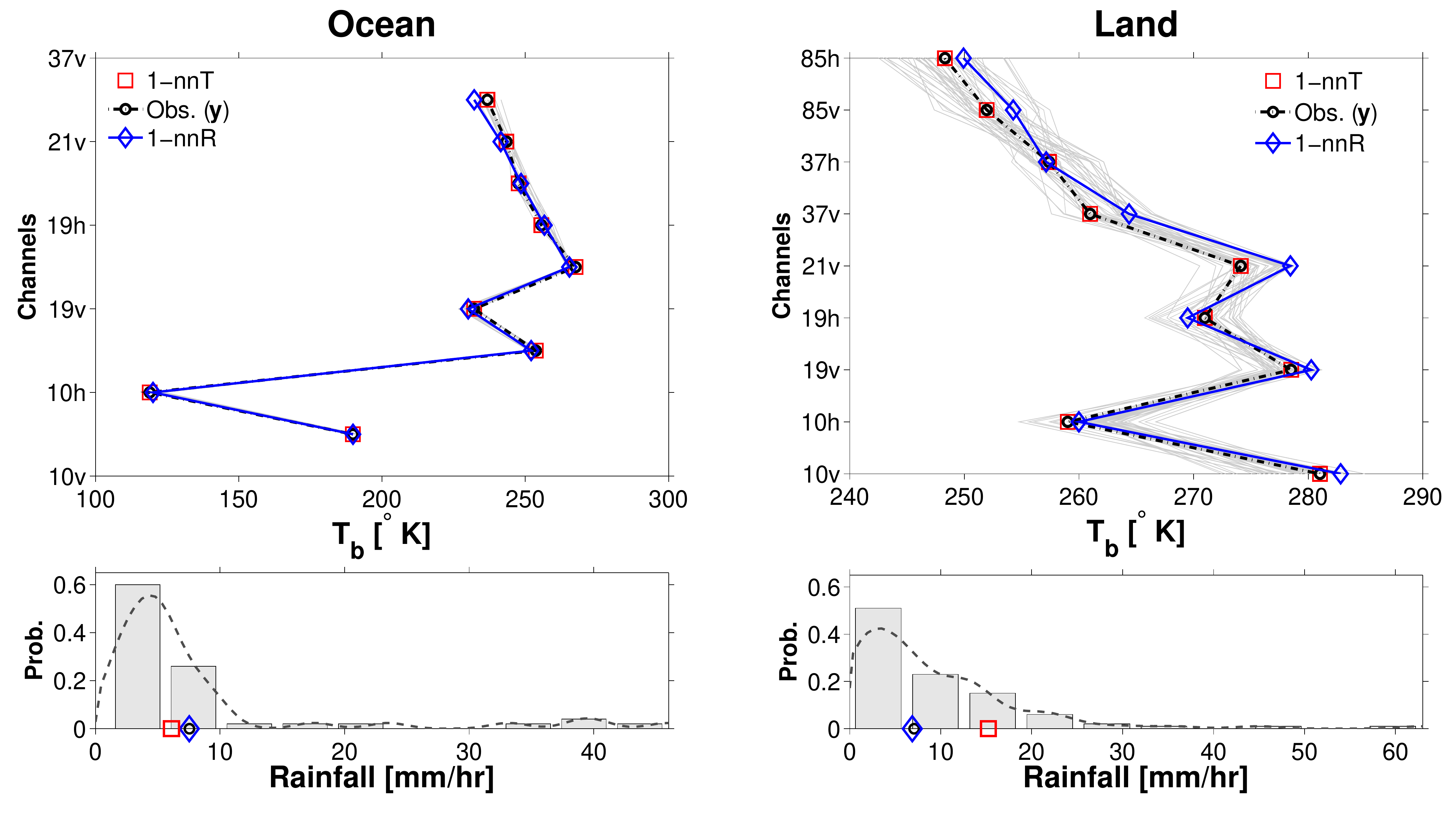} 
\par\end{centering}

\protect\caption{Top panels: Two arbitrary sampled raining vectors of the TMI-IB11
spectral brightness temperatures (dotted black lines with circles)
over ocean (left) and land (right). The gray lines are the 50-nearest
spectral neighbors in the Euclidean sense, obtained from an independent
learning set of the TMI-1B11 observations. Bottom panels: PR-2A25
surface rainfall probability histograms of the 50 spectral neighbors,
shown in the top panels. In top and bottom panels, the red squares
and the blue solid lines with diamonds show the 1-nearest neighbor
in the spectral (1-nnT) and rainfall (1-nnR) spaces, respectively.
\label{fig:2}}
\end{figure}
Furthermore, to better understand the correspondence between the neighboring
raining spectral brightness temperatures, in the Euclidean sense,
and their surface rainfall intensities, we independently collected
two learning sets of the form $\mathcal{L}=\left\{ \left(\mathbf{b}_{i},\, r_{i}\right)\right\} _{i=1}^{M}$
over ocean and land. Each set contains $M\approxeq10^{6}$ of coincidental
1B11 spectral brightness temperatures $\mathbf{b}\in\mathfrak{R}^{9}$
and their corresponding 2A25 surface rainfall $r\in\mathfrak{R}$
estimates. From a mathematical stand point, a simple nearest neighborhood
search reveals that the spectral temperatures over ocean and land
are not uniquely related to the estimated surface rainfall intensities
in the Euclidean sense \citep[see,][ for more discussion]{Lecuyer2002}.
Nevertheless, in the known lack of uniqueness, a basic question arises:
How can we obtain ``stable'' estimates of the surface rainfall using
neighboring spectral brightness temperatures in a properly collected
learning set? To this end, let us assume that a spectral vector of
brightness temperature is denoted by $\mathbf{y}$ and its scalar
surface rainfall value of interest is $x$. Top panels from left to
right in Fig.~\ref{fig:2} demonstrate two arbitrary vectors of the
1B11 raining brightness temperatures $\mathbf{y}\in\mathfrak{R}^{9}$
(black dashed lines) over ocean and land, together with their fifty
nearest neighbors $\left\{ \mathbf{b}_{k}\left(\mathbf{y}\right)\right\} _{k=1}^{K=50}$
(gray solid lines) obtained from the collected learning sets. Bottom
panels show the corresponding surface rainfall values $\left\{ r_{k}\left(\mathbf{y}\right)\right\} _{k=1}^{K=50}$
and their probability histograms. It turns out that all of the fifty
nearest spectral brightness temperatures were raining except for only
one of them over land. This observation implies that a supervised
nearest neighborhood classification, using coincidental TMI and PR
data, might be a very powerful approach for the rain/no-rain discrimination
problem. Furthermore, it can be seen that the first nearest neighbor
in the spectral space (1-nnT) does not necessary relate to the nearest
neighbor (1-nnR) in the rainfall space. However, in both cases, the
surface rainfalls of the neighboring spectral vectors are bounding
the rainfall values $x$ of interest. These bounds, both in the spectral
and rainfall spaces, are clearly tighter over ocean than over land,
mainly due to the stronger signal to noise ratio of the rainfall signatures.
Therefore, for each sampled $\mathbf{y}$, it can be naturally concluded
that a properly chosen statistic of $\left\{ r_{k}\left(\mathbf{y}\right)\right\} _{k=1}^{K}$
in the following form

\begin{equation}
\hat{x}=\sum_{k=1}^{K}c_{k}\, r_{k}\left(\mathbf{y}\right),\label{eq:1}
\end{equation}

\noindent may be adopted as a stable estimator of $x$, where $c_{k}$ denotes
some optimal weighting coefficients.

\section{SHRUNKEN LOCALLY LINEAR EMBEDDING FOR RETRIEVAL OF PRECIPITATION\label{sec:3}}

\subsection{Rainfall Retrieval as an Inverse Problem}

Passive rainfall retrieval in the microwave bands can be considered
as a nonlinear inverse problem, where its solution shall be constrained
by the underlying laws of atmospheric thermal radiative transfer in
a weak or strong sense \citep[e.g.,][]{Jan94,Sar03}. To recast the
microwave rainfall retrieval in a standard form of a discrete inverse
problem, let us assume that each vector of spectral brightness temperatures
and their corresponding rainfall profiles are $\mathbf{y}=\left(y_{1},y_{2},\ldots,y_{n_{c}}\right)^{{\rm T}}$
and $\mathbf{x}=\left(x_{1},x_{2},\ldots,x_{n_{r}}\right)^{{\rm T}}$,
respectively, where $n_{c}$ and $n_{r}$ denote the number of spectral
channels and vertical layers of the rainfall intensity profile. As
a result, in a finite dimension, spectral observations might be related
to the rainfall intensity profile through the following nonlinear
observation model:

\begin{equation}
\mathbf{y}=\mathcal{F}\left(\mathbf{x}\right)+\mathbf{v},\label{eq:2}
\end{equation}

\noindent where, $\mathcal{F}\left(\cdot\right):\,\mathbf{x}\rightarrow\mathbf{y}$
can be considered to be a functional representation of the radiative
transfer equations that maps the rainfall intensity profiles onto
the space of spectral brightness temperatures, and $\mathbf{v}\in\mathfrak{R}^{n_{c}}$
represents the observation error with a finite energy. Obviously,
the goal of the retrieval is to obtain an estimate of the rainfall
profile $\mathbf{x}$, given spectral brightness temperatures $\mathbf{y}$,
the radiative transfer functional $\mathcal{F}\left(\cdot\right)$,
and a priori information about the error. The search for a stable
closed form solution of the above inverse problem seems almost a hopeless
quest at least for now, given the fact that $\mathcal{F}\left(\cdot\right)$
is extremely nonlinear, especially under the scattering dominant regime.
In the subsequent section it will be clear that our algorithm is indeed
a workaround to this complex inverse problem.

\subsection{Algorithm}

Motivated by our observations in Section \ref{sec: 2}, to bridge
the explained complexities in the rainfall retrieval problem, our
algorithm relies on a priori collected database or say learning set
denoted by $\mathcal{L}=\left\{ \left(\mathbf{b}_{i},\,\mathbf{r}_{i}\right)\right\} _{i=1}^{M}$.
This set is populated by a large number of coincidental brightness
temperatures $\mathbf{b}_{i}=\left[b_{1i},b_{2i},\ldots,b_{n_{c}i}\right]^{{\rm T}}\in\mathfrak{R}^{n_{c}}$
and their corresponding rainfall profiles $\mathbf{r}_{i}=\left[r_{1i},r_{2i},\ldots,r_{n_{r}i}\right]^{{\rm T}}\in\mathfrak{R}^{n_{r}}$.
For notational convenience, let us stack these pairs according to
a fixed order in two joint matrices $\mathbf{B}=\left[\mathbf{b}_{1}|\ldots|\mathbf{b}_{M}\right]\in\mathfrak{R}^{n_{c}\times M}$
and $\mathbf{R}=\left[\mathbf{r}_{1}|\ldots|\mathbf{r}_{M}\right]\in\mathfrak{R}^{n_{r}\times M}$,
which are called the spectral and rainfall ``dictionaries''. In our
notation, each of these pairs are called elementary ``atoms'' to
be used for reconstruction of the rainfall fields from their observed
spectral signatures. As is evident, theses dictionaries can be populated
either by observational or physically-based generated pairs.

In the detection step, we simply use a supervised nearest neighborhood
classification rule, guided by the dictionaries. In particular, for
a given observation vector of spectral brightness temperature $\mathbf{y}\in\mathfrak{R}^{n_{c}}$
and the dictionary pair $\left(\mathbf{B},\,\mathbf{R}\right)$, let
us assume that $\mathcal{S}$ denotes the set of $K$ column indices
of $\mathbf{B}$ that contain the nearest spectral atoms to $\mathbf{y}$
in the Euclidean sense. Given this set, the algorithm forms two joint
sub-dictionaries $\left(\mathbf{B}_{\mathcal{S}}\in\mathfrak{R}^{n_{c}\times K},\:\mathbf{R}_{\mathcal{S}}\in\mathfrak{R}^{n_{r}\times K}\right)$,
which are generated by those $K=\left|\mathcal{S}\right|$ nearest
spectral $\left\{ \mathbf{b}_{k}\right\} _{k=1}^{K}\in\mathbf{B}$
and their corresponding rainfall atoms $\left\{ \mathbf{r}_{k}\right\} _{k=1}^{K}\in\mathbf{R}$.
Assuming that the last row of the rainfall sub-dictionary $\mathbf{R}_{\mathcal{S}}$
contains the near surface rainfall intensity values, the algorithm
simply makes use of a probabilistic voting to declare $\mathbf{y}$
as raining or non-raining. In other words, choosing a probability
threshold $p$, the algorithm labels $\mathbf{y}$ as raining, if
more than $p\, K$ number of $\left\{ \mathbf{r}_{k}\right\} _{k=1}^{K}$
are raining at the surface. In the estimation step, motivated by the
results in Fig.~\ref{fig:2}, we assume that the true rainfall profile
$\mathbf{x}$ of the given spectral observation $\mathbf{y}$ can
be well explained by the $\mathbf{R}_{\mathcal{S}}$'s atoms, through
the following linear model:

\begin{equation}
\mathbf{x}=\mathbf{R}_{\mathcal{S}}\mathbf{c}+\mathbf{e},\label{eq:3}
\end{equation}

where $\mathbf{c}\in\mathfrak{R}^{K}$ is a vector of representation
coefficients that linearly combines atoms of the rainfall sub-dictionary
and $\mathbf{e}\in\mathfrak{R}^{n_{r}}$ denotes a zero mean error
with finite energy. As a result, given an estimate of the representation
coefficients $\hat{\mathbf{c}}$, conditional expectation of the rainfall
profile $\hat{\mathbf{x}}$ can be obtained as follows:

\begin{equation}
\hat{\mathbf{x}}=\mathbb{E}\left(\mathbf{x}|\hat{\mathbf{c}}\right)=\mathbf{R}_{\mathcal{S}}\hat{\mathbf{c}}.\label{eq:4}
\end{equation}

Obviously, estimation of the representation coefficients solely from
equation (\ref{eq:3}) is ambiguous as both sides of the equation
are unknown. To find a solution, as previously explained, we assume
that the neighboring rainfall profiles and their spectral signatures
live close to two smooth manifolds with analogous geometric structure
and thus similar locally linear representation. Therefore, the algorithm
assumes a spectral observation model with the same linear representation
coefficients as follows:

\begin{equation}
\mathbf{y}=\mathbf{B}_{\mathcal{S}}\mathbf{c}+\mathbf{v},\label{eq:5}
\end{equation}

\noindent where $\mathbf{v}\in\mathfrak{R}^{n_{c}}$ denotes a zero mean error
with finite energy. As is evident, estimation of the representation
coefficients from (\ref{eq:5}) is no longer an ill-defined problem.
To estimate the representation coefficients in this linear model,
the weighted Minimum Mean Squared Error (MMSE) estimator, constrained
to the probability simplex, seems to be the first choice as follows:
\begin{equation}
\begin{aligned} & \underset{\mathbf{c}}{\text{minimize}} &  & \left\Vert \mathbf{W}^{1/2}\left(\mathbf{y}-\mathbf{B}_{\mathcal{S}}\mathbf{c}\right)\right\Vert _{2}^{2}\\
 & \text{subject to} &  & \mathbf{c}\succeq0,\enskip\mathbf{1}^{{\rm T}}\mathbf{c}=1,
\end{aligned}
\label{eq:6}
\end{equation}

\noindent where the $\ell_{2}$-norm is $\left\Vert \mathbf{c}\right\Vert _{2}^{2}=\Sigma_{i}c_{i}^{2}$,
$\mathbf{c}\succeq0$ implies element-wise non-negativity and the
positive definite $\mathbf{W}\succ0$ in $\mathfrak{R}^{n_{c}\times n_{c}}$
determines the relative importance or weights of each channel. These
weights may be chosen to relatively encode the signal-to-noise ratio
of the spectral raining signatures. Note that the non-negativity constraint
is required to be physically consistent with the positivity of the
brightness temperatures in Kelvin. Furthermore, the sum to one constraint
assures that the estimates are locally unbiased. More importantly,
this equality constraint makes the solution invariant to rotation,
rescaling, and translation of the neighboring spectral observations
\citep[see, ][]{RowSL00}. For similar concept in rainfall downscaling,
the reader is also referred to \citep{EbtF12a} and \citep{EfgEZa13}.

However, problem (\ref{eq:6}) is likely to be severely ill-posed
due to the observation noise, especially when the column dimension
of $\mathbf{B}_{\mathcal{S}}$ is larger than that of spectral bands
$n_{c}$. To make the problem well-posed and sufficiently stable,
we suggest the following regularization scheme

\begin{equation}
\begin{aligned} & \underset{\mathbf{c}}{\text{minimize}} &  & \left\Vert \mathbf{W}^{1/2}\left(\mathbf{y}-\mathbf{B}_{\mathcal{S}}\mathbf{c}\right)\right\Vert _{2}^{2}+\lambda_{1}\left\Vert \mathbf{c}\right\Vert _{1}+\lambda_{2}\left\Vert \mathbf{c}\right\Vert _{2}^{2}\\
 & \text{subject to} &  & \mathbf{c}\succeq0,\enskip\mathbf{1}^{{\rm T}}\mathbf{c}=1,
\end{aligned}
\label{eq:7}
\end{equation}

\noindent where the $\ell_{1}$-norm is $\left\Vert \mathbf{c}\right\Vert _{1}=\Sigma_{i}\left|c_{i}\right|$,
$\lambda_{1},\,\lambda_{2}$ are non-negative regularization parameters.
Obviously, obtaining $\hat{\mathbf{c}}$ as the solution of the above
problem, we can retrieve the rainfall using expression (\ref{eq:4})
as $\hat{\mathbf{x}}=\mathbf{R}_{\mathcal{S}}\hat{\mathbf{c}}$. 

{\small{}}
\begin{algorithm}[t]
\textbf{\small{}Input: }{\small{}Spectral observations $\mathbf{Y}$
containing $\left\{ \mathbf{y}_{i}=\left[y_{1i},y_{2i},\ldots,y_{n_{c}i}\right]^{{\rm T}}\in\mathfrak{R}^{n_{c}}\right\} _{i=1}^{N}$
vectors of spectral brightness temperatures, spectral $\mathbf{B}\in\mathfrak{R}^{n_{c}\times M}$and
rainfall $\mathbf{R}\in\mathfrak{R}^{n_{r}\times M}$ dictionaries,
weight matrix $\mathbf{W}\in\mathfrak{R}^{n_{c}\times n_{c}}$, detection
probability $p$, number of nearest neighborhoods $K$, and regularization
parameters $\lambda_{1}$, $\lambda_{2}$.}{\small \par}

\textbf{\small{}Output:}{\small{} Precipitation field $\mathbf{X}$
containing $\left\{ \mathbf{x}_{i}\in\mathfrak{R}^{n_{r}}\right\} _{i=1}^{N}$
pixels of rainfall intensity profiles.}{\small \par}

\textbf{\small{}For }{\small{}$i:=1$ }\textbf{\small{}to}{\small{}
$N$ step 1 }\textbf{\small{}do}{\small{} }{\small \par}
\begin{itemize}
\item {\small{}Find sub-dictionaries $\mathbf{B}_{\mathcal{S}}\in\mathfrak{R}^{n_{c}\times K}$
and $\mathbf{R}_{\mathcal{S}}\in\mathfrak{R}^{n_{r}\times K}$, where
$\mathcal{S}$ is the set of column indices of $\mathbf{B}$ which
contains the $k$-nearest neighborhoods of $\mathbf{y}_{i}$.}{\small \par}
\item {\small{}Let $\mathbf{R}_{\mathcal{S}}\left(\text{end},\,:\right)$
denotes the last row of $\mathbf{R}_{\mathcal{S}}$ containing neighboring
surface rainfall.}{\small \par}
\item \textbf{\small{}If}{\small{} $\left|{\rm supp}\left(\mathbf{R}_{\mathcal{S}}\left(\text{end},\,:\right)\right)\right|\geq p\, K$
, }\\
{\small{}}\\
{\small{}$\;$- Standardize $\mathbf{y}_{i}$ and atoms of $\mathbf{B}_{\mathcal{S}}$,
such that $\sum_{j}^{n_{c}}y_{ji}=0$ , $\sum_{j}^{n_{c}}b_{jk}=0$
, and $\sum_{j}^{n_{c}}b_{jk}^{2}=1$ , for $k=1,\ldots,K$. }\\
{\small{}$\;$- $\hat{\mathbf{c}}_{i}={\rm argmin}_{\mathbf{c}_{i}\succeq0,\:\mathbf{1}^{{\rm T}}\mathbf{c}_{i}=1}\left\{ \left\Vert \mathbf{W}^{1/2}\left(\mathbf{y}_{i}-\mathbf{B}_{\mathcal{S}}\mathbf{c}_{i}\right)\right\Vert _{2}^{2}+\lambda_{1}\left\Vert \mathbf{c}_{i}\right\Vert _{1}+\lambda_{2}\left\Vert \mathbf{c}_{i}\right\Vert _{2}^{2}\right\} $}\\
{\small{}$\;$- $\hat{\mathbf{x}}_{i}=\mathbf{R}_{\mathcal{S}}\hat{\mathbf{c}}_{i}$}\\
{\small{}}\\
{\small{} }\textbf{\small{}else}{\small{}}\\
{\small{}}\\
{\small{}$\;$- $\hat{\mathbf{x}}_{i}=0$}\\
{\small{}}\\
{\small{} }\textbf{\small{}End If}{\small{} }{\small \par}
\end{itemize}
\textbf{\small{}End For}{\small \par}

{\small{}\protect\caption{\textbf{\small{}Sh}{\small{}runken Locally Linear Embedding }\textbf{\small{}A}{\small{}lgorithm
for }\textbf{\small{}R}{\small{}etrieval of }\textbf{\small{}P}{\small{}recipitation
(ShARP).} \label{alg:1}}
}
\end{algorithm}
{\small \par}

\begin{figure}[t]
\noindent \begin{centering}
\includegraphics[scale=0.7]{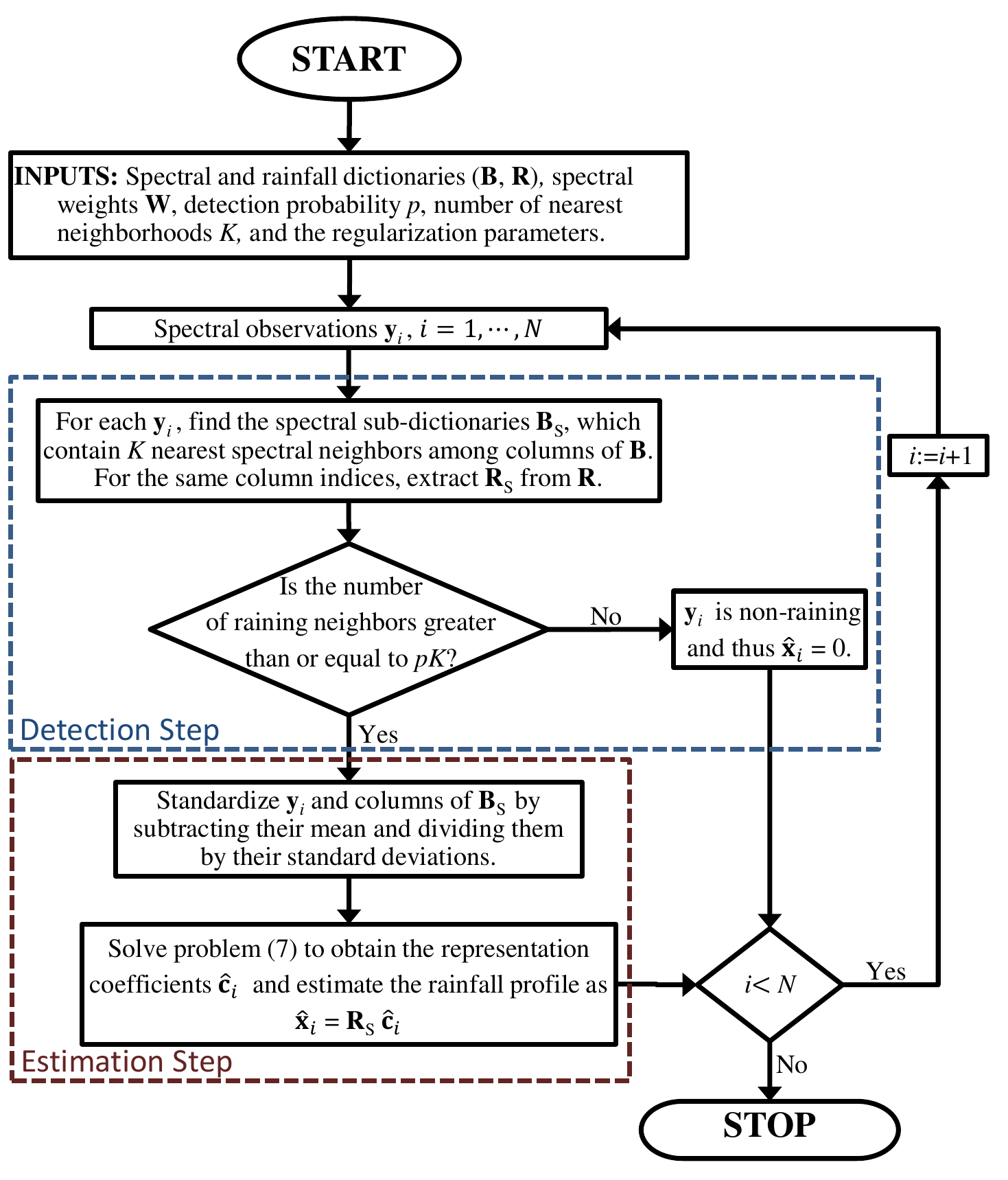}
\par\end{centering}

\protect\caption{Flowchart of the ShARP algorithm. See Algorithm \ref{alg:1} for more
detailed explanation. \label{fig:3}}
\end{figure}

Note that problem (\ref{eq:7}) is a non-smooth convex problem. It
is non-smooth as the $\ell_{1}$-norm is not differentiable at the
origin. Convexity arises as it uses a conic combination of two well-known
convex penalty functions to regularize a classic weighted least-squares
problem over a convex set. These two regularization functions have
been widely used to properly narrow down the solution of ill-posed
inverse problems. In under-determined system of equations, the $\ell_{1}$-norm
penalty has proven to be an effective regularization for obtaining
``sparse'' solutions. In other words, it turns out that this regularization
promotes sparsity in the solutions, as it uses a minimal number of
atoms of $\mathbf{B}_{\mathcal{S}}$, while retains maximum amount
of information \citep{Don95,Tib96,CheDS98,Ela10}. On the other hand,
the $\ell_{2}$-norm penalty is the most widely used regularization
approach to stabilize the solutions of ``dense'' ill-posed inverse
problems while it incorporates all the atoms of $\mathbf{B}_{\mathcal{S}}$
\citep{Tik77,Han10} in the solution. Confining the regularization
in (\ref{eq:7}) solely to the $\ell_{1}$-norm ($\lambda_{2}=0$)
is restrictive for rainfall retrieval in the current setting of our
algorithm because of two main reasons. First, the number of selected
columns of $\mathbf{B}_{\mathcal{S}}$, or say non-zero elements of
the representation coefficients, will be bounded in this case by the
number of the available spectral bands $n_{c}$. Second, the spectral
atoms in the sub-dictionary $\mathbf{B}_{\mathcal{S}}$ are likely
to be highly correlated and clustered in groups. In this condition,
the $\ell_{1}$-norm regularization typically fails to take into account
the contribution of clustered atoms. On the other hand, all of the
spectral atoms in $\mathbf{B}_{\mathcal{S}}$ will be taken into account
if we solely rely on the $\ell_{2}$ penalty, which can lead to selection
of irrelevant atoms and overly smooth rainfall retrieval. However,
the proposed mixed penalty removes the explained limitations of each
individual regularization scheme through stabilizing the problem regularization
path, encouraging grouping effects by shrinking the clusters of correlated
atoms and averaging their representation coefficients \citep[see,][]{ZouH05}.
In addition, from a practical point of view, this mixed regularization
increases the flexibility of the algorithm to cope with the ill-conditioning
arising due to the presence of very similar and correlated atoms in
the spectral sub-dictionary. This property is extremely desirable
especially for the future developments of our algorithm to accommodate
both observationally and physically generated dictionaries. Throughout
this paper, we consider a convex combination of regularization penalty
functions by assuming $\lambda_{2}=\lambda\alpha$ and $\lambda_{1}=\lambda\left(1-\alpha\right)$
for all $\alpha\in\left(0,\,1\right)$. As we use the concept of locally
linear embedding together with the above mixed shrinkage estimation,
we call our retrieval technique the \textbf{Sh}runken Locally Linear
Embedding \textbf{A}lgorithm for \textbf{R}etrieval of \textbf{P}recipitation
(ShARP). The details are summarized in Algorithm \ref{alg:1} and
sketched in Fig.~\ref{fig:3}. Given the induced non-negativity constraint
in problem (\ref{eq:7}) allows us to solve it via a constrained quadratic
programming (QP) as follows:

\begin{equation}
\begin{aligned} & \underset{\mathbf{c}}{\text{minimize}} &  & \mathbf{c}^{{\rm T}}\left(\mathbf{B}_{\mathcal{S}}^{{\rm T}}\mathbf{W}\mathbf{B}_{\mathcal{S}}+\lambda_{2}\mathbf{I}\right)\mathbf{c}+\left(\lambda_{1}\mathbf{1}-\mathbf{B}_{\mathcal{S}}^{{\rm T}}\mathbf{W}^{{\rm T}}\mathbf{y}\right)^{{\rm T}}\mathbf{c}\\
 & \text{subject to} &  & \mathbf{c}\succeq0,\enskip\mathbf{1}^{{\rm T}}\mathbf{c}=1,
\end{aligned}
\label{eq:8}
\end{equation}

\noindent where $\mathbf{1}=\left[1,\ldots,1\right]^{{\rm T}}\in\mathfrak{R}^{K}$.

It is important to note that problem (\ref{eq:7}) is, in effect,
a constrained Bayesian Maximum a Posteriori (MAP) estimator under
the following prior 

\begin{equation}
p\left(\mathbf{c}\right)\propto\exp\left(-\lambda_{1}\left\Vert \mathbf{c}\right\Vert _{1}-\lambda_{2}\left\Vert \mathbf{c}\right\Vert _{2}^{2}\right),
\end{equation}

which is a conic combination of the Gaussian and Laplace densities
\citep[see,][]{ZhoEM05}. Therefore, the posterior density of the
estimated coefficients and thus rainfall values is not Gaussian. As
a result, closed form uncertainty analysis of the retrieved rainfall
is not trivial and may be addressed through randomization or ensemble
analysis. To this end, one can simply see that the rows of the sub-dictionary
$\mathbf{R}_{\mathcal{S}}$ contain $K$-samples of the posterior
Probability Density Function (PDF) of the neighboring rainfall intensity
profiles. Thus, depending on the selected number of the nearest neighbors,
the whole posterior PDF of the ShARP estimator can be empirically
approximated by counting the relative frequency of the rainfall occurrence.
This strategy will be used in the sequel to estimate the uncertainty
of the retrieved rainfall. 

\section{EXPERIMENTS USING TRMM DATA\label{sec:4}}

As previously explained, in the current implementation of ShARP, we
only confine our consideration to empirical rainfall and spectral
dictionaries collected from the coincidental PR-2A25 and TMI-1B11
products and only retrieve surface rainfall. Therefore, the 2A25 product
can be used as a reference to validate the results of the ShARP algorithm.
To further examine the pros and cons of its performance, all of the
retrieval experiments are also shown versus the surface rainfall obtained
from the standard passive TMI-2A12 retrieval product.

\subsection{Algorithm Setup}

\begin{figure}[H]
\noindent \begin{centering}
\includegraphics[width=0.77\paperwidth]{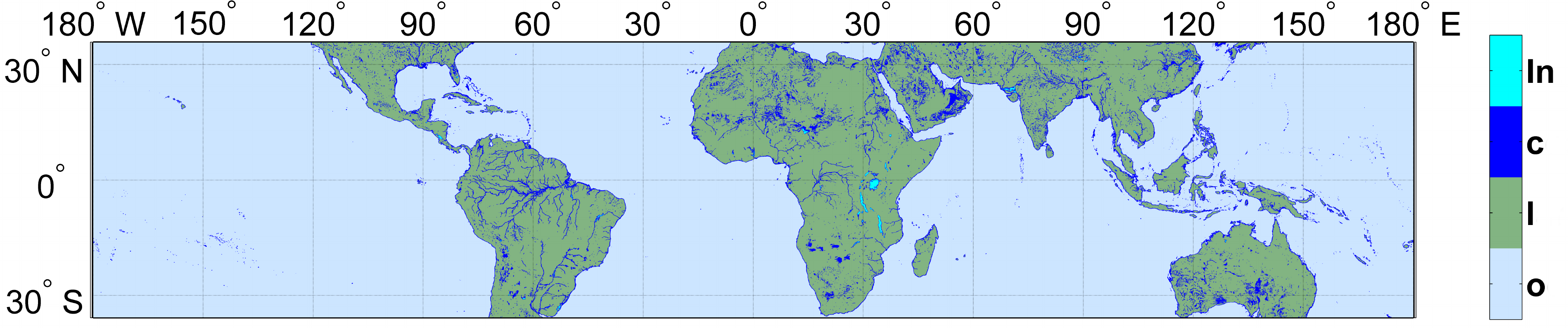}
\par\end{centering}

\protect\caption{Different earth surface classes used in the current version of the
ShARP, namely inland water body (\textbf{In}), coastal zone (\textbf{c}),
land (\textbf{l}) and ocean (\textbf{o}). The classification is adopted
based on the available data (version 7) of the PR-1C21 product, which
are mapped onto a $0.05$-degree regular grid. \label{fig:4} }
\end{figure}

In the current implementation of ShARP, we defined four different
earth surface classes, namely: ocean, land, coast and inland water
(Fig.~\ref{fig:4}). In other words, we collected four dictionaries
over each surface class and use them in Algorithm~\ref{alg:1} depending
on the geolocation of a given pixel of the observed spectral brightness
temperatures. This surface stratification is obtained from standard
surface data in the PR-1C21 product (version 7) at $\sim5\times5$
km grid box. In this classification, the coastal areas are referred
to those locations on the globe, where the presence of water is not
permanent due to the seasonal variations or tidal effects. To construct
spectral and rainfall dictionaries, we randomly sampled 750 orbits
from our rainfall database. In these sampled orbits, more than $25\times10^{6}$
pairs of raining and non-raining signatures were used to construct
the required dictionaries.

\subsubsection{Detection Step}

As previously explained, rain/no-rain classification from microwave
observations and its induced error on the quality of rainfall retrieval
has been addressed in numerous studies \citep[e.g.,][]{Gro91,KumG94a,LiB96,FerETAL98,Seto05},
and reported as a challenging problem which is not easy to mitigate,
especially over land \citep[see,][]{KumRCRB10}. Therefore, in developing
rainfall retrieval techniques, we naturally have a choice to either
first detect the storm raining areas and then estimate the rainfall
intensities or just use an estimation scheme that automatically recovers
the raining areas. In general, rainfall retrieval with a sequential
rain/no-rain detection and estimation scheme may be advantageous in
the sense that it allows us to control the probability of false alarm
while confining the computational expense of estimation only to the
detected raining areas.

\begin{wrapfigure}{c}{0.5\textwidth}
\noindent \begin{centering}
\includegraphics[width=0.31\textwidth]{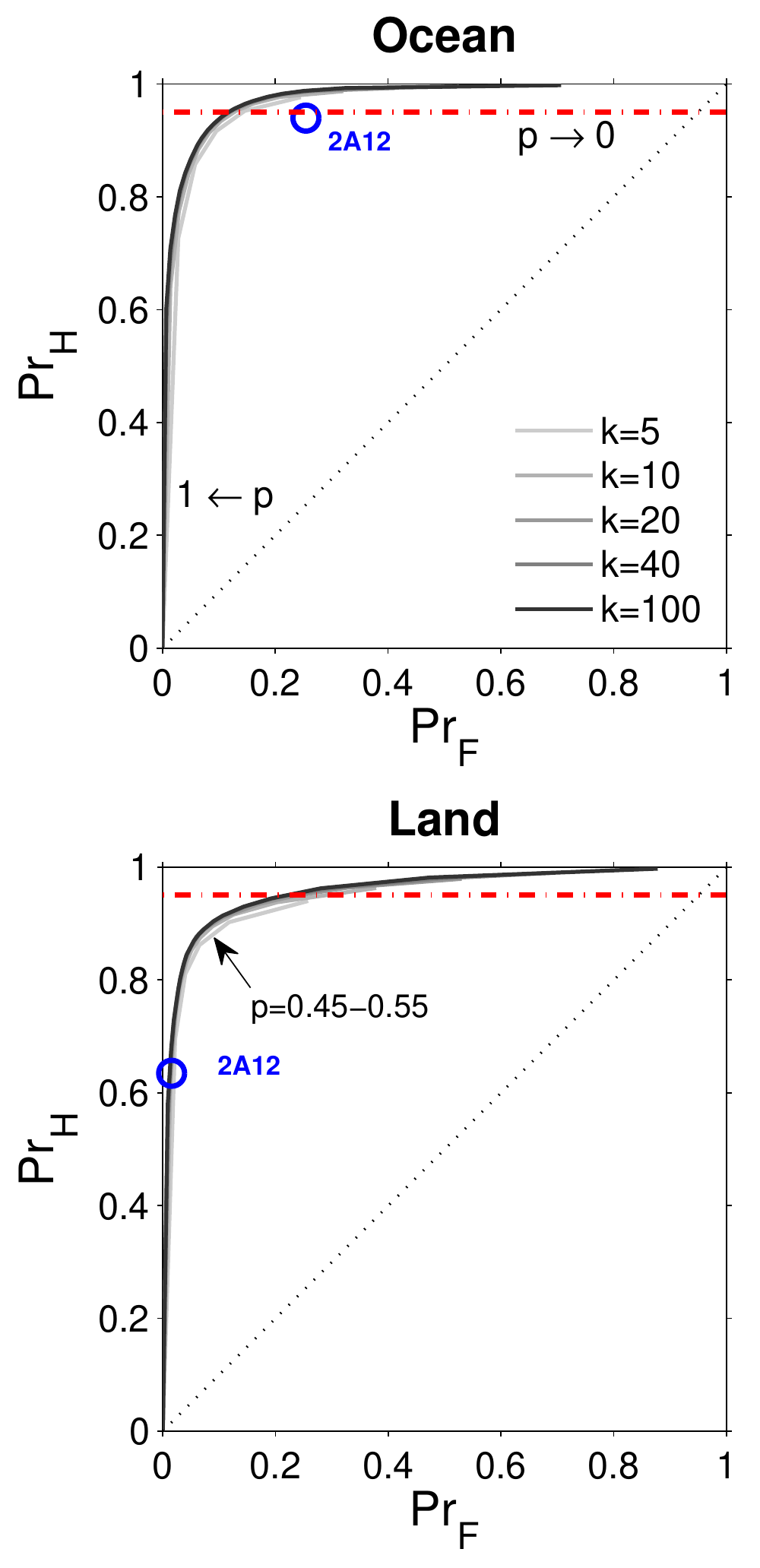}
\par\end{centering}

\protect\caption{Rainfall Receiver Operating Characteristic (ROC) curve over ocean
top panel) and land (bottom panel) for different probability of detection
$p\in[0,\,1]$ and number of nearest neighborhoods $K\in\left\{ 5,\,10,\,20,\,40,\,100\right\} $
of the ShARP algorithm. The blue circles show the 2A12 (version 7)
product and the red dash-dotted lines show the 0.95 probability of
hit as a datum. \label{fig:5}}
\end{wrapfigure}

Considering 2A25 as a reference rainfall field to validate ShARP,
Fig.~\ref{fig:5} shows the estimated probability of hit (${\rm Pr}_{{\rm H}}$)
versus probability of false alarm (${\rm Pr}_{{\rm F}}$) for the
rain/no-rain detection step of our algorithm as the classification
parameters are varied. Here, the results are obtained by applying
the detection step to more than $3\times10^{5}$ randomly chosen pixels
of spectral observations from our rainfall database. Note that, these
spectral pixels are selected randomly from our rainfall database and
have not been used in the construction of the retrieval dictionaries.
In Fig.~\ref{fig:5}, we can see that the ShARP classification rule
is not very sensitive to the number of chosen nearest neighborhoods
as all of the curves are nearly collapsing onto each other. The ShARP
rain/no-rain detection quality for $K=20$ and the majority vote rule,
that is $p=0.5$, is presented in Table~\ref{tab:1}. This table
explains that over ocean and land, our algorithm matches raining pixels
of the 2A25 in 96 and 90\% of the cases while the false alarm rate
does not exceed 8\% and 6\%, respectively. Fig.~\ref{fig:5} also
shows the position of the 2A12 retrieval product. It is seen that,
given the 2A25 is raining over ocean, the 2A12 is raining in 95\%
of the cases. On the other hand, we see that in 20\% of the cases
2A12 detects raining areas which may have been missed by the 2A25
and thus ShARP. Although interpretation of this discrepancy is not
central to the thrust of this paper, this result seems to be consistent
with the recent evidence from the CloudSat satellite suggesting that
the PR underestimates the extent of light rain over ocean \citep{EllT09},
which may reach up to 10\% of the rainfall volume on average over
the tropics \citep{Berg2009}. Conversely, over land, we see that
if 2A25 is raining, ShARP is raining in 90\% of cases while 38\% of
these raining pixels are not captured in the 2A12.


\begin{table}[h]
\begin{centering}
{\scriptsize{}}%
\begin{tabular}{c|c|>{\centering}p{9mm}|>{\centering}p{10mm}|>{\centering}p{9mm}|>{\centering}p{10mm}|}
\multicolumn{2}{c}{} & \multicolumn{4}{c}{\textbf{\scriptsize{}Observation (2A25)}}\tabularnewline
\cline{3-6} 
\multicolumn{2}{c|}{} & \multicolumn{2}{c|}{{\scriptsize{}Ocean}} & \multicolumn{2}{c|}{{\scriptsize{}Land}}\tabularnewline
\cline{3-6} 
\multicolumn{2}{c|}{} & {\scriptsize{}rain} & {\scriptsize{}no-rain} & {\scriptsize{}rain} & {\scriptsize{}no-rain}\tabularnewline
\cline{2-6} 
\multirow{2}{*}{\textbf{\scriptsize{}Detection (ShARP)}} & {\scriptsize{}rain} & {\scriptsize{}0.96} & {\scriptsize{}0.08} & {\scriptsize{}0.90} & {\scriptsize{}0.06}\tabularnewline
\cline{2-6} 
 & {\scriptsize{}no-rain} & {\scriptsize{}0.04} & {\scriptsize{}0.92} & {\scriptsize{}0.1} & {\scriptsize{}0.94}\tabularnewline
\cline{2-6} 
\end{tabular}
\par\end{centering}{\scriptsize \par}

\protect\caption{Probability of hit and false alarm for twenty nearest neighbors $K=20$
and probability threshold of $p=0.5$. The results are obtained by
comparing ShARP with 2A25. \label{tab:1}}
\end{table}

\subsubsection{Estimation Step}

After finding the storm raining areas, our algorithm moves toward
estimation of the rainfall intensities. Recall that, we use a positive
definite weight matrix $\mathbf{W}$ in problem (\ref{eq:7}) that
determines the relative importance of each channel over different
surface classes. To design this weight matrix, we use the normalized
coefficients of variation for each channel as reported in Fig.~\ref{fig:1}.
In particular, the relative weight of the $i^{\text{th}}$ channel
for a specific surface class is obtained by normalizing its coefficient
of variation as $w_{i}=c_{v}^{i}/\underset{i}{\max}\left(c_{v}^{i}\right)$,
$i=1,\,\ldots,\,9$ (Table~\ref{tab:2}). The weight matrix is then
assigned to be $\mathbf{W}=\text{diag}\left(w_{i}\right)$. Using
these weights allows us to make the least-squares term in problem
(\ref{eq:7}) invariant to temperature translations among spectral
channels and more responsive to stronger rainfall signal-to-noise
ratio. In other words, these weights reduce saturation of the cost
due to some excessively cold and/or warm channels while maintain it
sensitive to their relative variability. To solve problem (\ref{eq:7}),
we use a primal-dual interior-point method \citep[see,][chap. 11]{BoyV04}.
Basically, in this class of convex optimization techniques, the inequality
constrained quadratic programing problem (\ref{eq:8}) is reformulated
into an equality constrained problem to which iterative Newton's method
can be applied. Specifically, we employed Linear-programing Interior
Point SOLver (lIPSOL) \citep{Zha95} which is based on a variant of
the algorithm by \citet{Meh92}. In this optimization sub-algorithm
the maximum number of iterations in Newton's steps is set to 200,
the termination tolerance on the function value and magnitude of relative
changes in the optimization variable are both set to $1\text{e}-8$.
We set the algorithm regularization parameters to be $\lambda=0.001$
and $\alpha=0.1$, which appears to work well for a wide range of
rainfall retrieval experiments. This setting permits the algorithm
to perform a full orbital rainfall retrieval in the order of 10 to
15 minutes on a contemporary desktop machine.

\begin{table}[h]
\noindent \begin{centering}
{\small{}}%
\begin{tabular}{|c|c|c|c|c|c|c|c|c|c|}
\hline 
\multicolumn{10}{|c|}{\textbf{\small{}Relative weights }}\tabularnewline
\hline 
\multirow{2}{*}{{\small{}Surface Classes}} & \multicolumn{9}{c|}{{\small{}Channels}}\tabularnewline
\cline{2-10} 
 & {\small{}10v}\textbf{\small{} } & {\small{}10h}\textbf{\small{} } & {\small{}19v}\textbf{\small{} } & {\small{}19h}\textbf{\small{} } & {\small{}21v}\textbf{\small{} } & {\small{}37h}\textbf{\small{} } & {\small{}37v}\textbf{\small{} } & {\small{}85v}\textbf{\small{} } & {\small{}85h}\tabularnewline
\hline 
\hline 
{\small{}Ocean } & {\small{}0.39} & {\small{}1.00 } & {\small{}0.35 } & {\small{}0.76} & {\small{}0.19} & {\small{}0.14} & {\small{}0.40 } & {\small{}0.49} & {\small{}0.45}\tabularnewline
\hline 
{\small{}Land } & {\small{}0.07 } & {\small{}0.17} & {\small{}0.09} & {\small{}0.09} & {\small{}0.12} & {\small{}0.37} & {\small{}0.35 } & {\small{}1.00} & {\small{}0.97}\tabularnewline
\hline 
{\small{}Coast } & {\small{}0.19} & {\small{}0.42} & {\small{}0.13 } & {\small{}0.36} & {\small{}0.07} & {\small{}0.26} & {\small{}0.20 } & {\small{}1.00} & {\small{}0.95}\tabularnewline
\hline 
{\small{}Inland-water } & {\small{}0.33 } & {\small{}0.66} & {\small{}0.36 } & {\small{}0.84 } & {\small{}0.20} & {\small{}0.26} & {\small{}0.59} & {\small{}1.00} & {\small{}0.88}\tabularnewline
\hline 
\end{tabular}
\par\end{centering}{\small \par}

\protect\caption{The diagonal elements of the weight matrix $\mathbf{W}\in\mathfrak{R}^{9\times9}$
used in the ShARP algorithm for the chosen earth surface classes.
\label{tab:2}}
\end{table}
\subsection{Instantaneous Retrieval Experiments }

\begin{figure}[h]
\noindent \begin{centering}
\includegraphics[width=0.74\paperwidth]{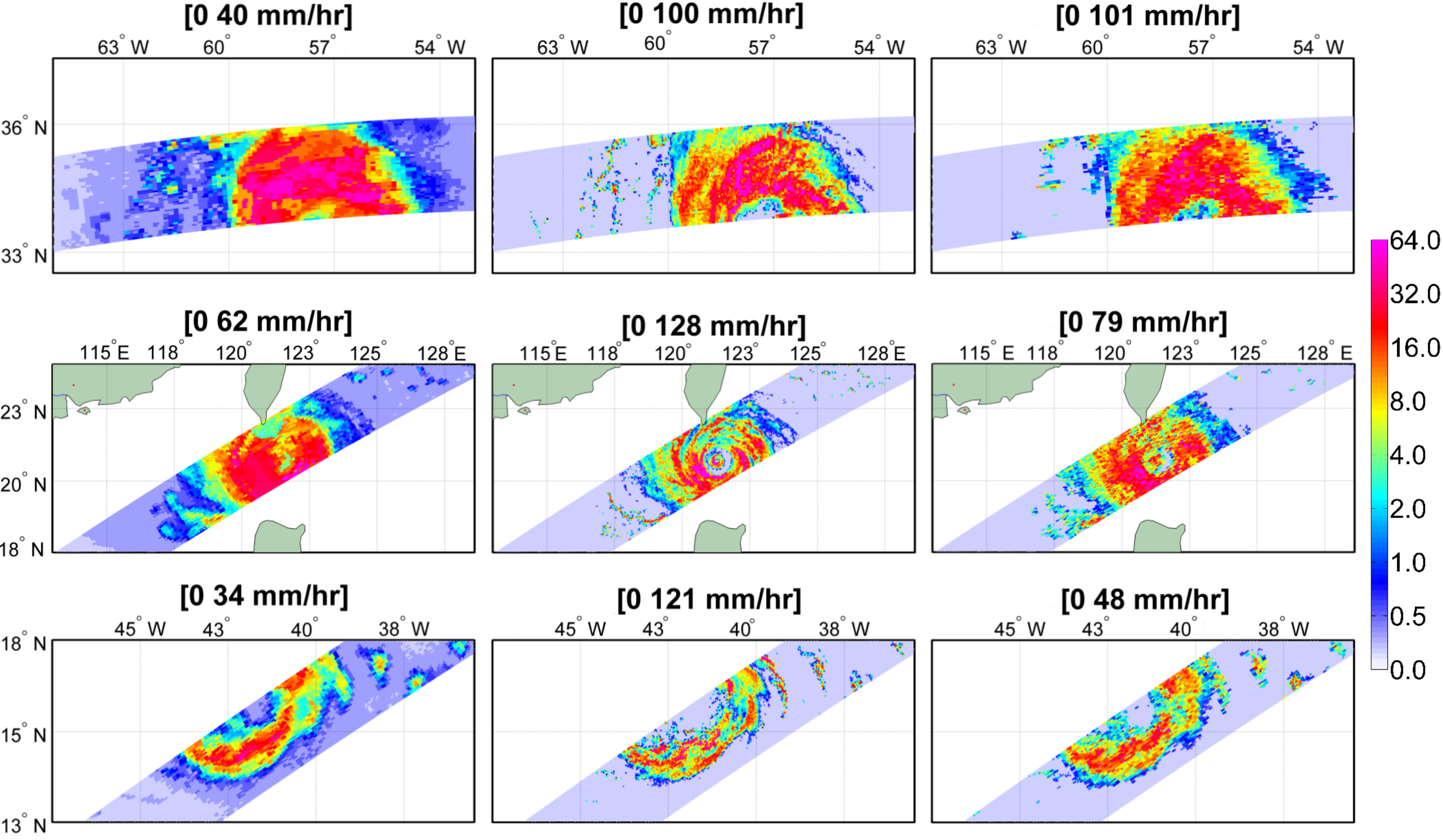}
\par\end{centering}

\protect\caption{From left to right: TMI-2A12, PR-2A25 and ShARP retrievals. Top to
bottom panels: hurricane Danielle in 08/29/2010 (orbit No. 72840)
at 09:48 UTC; super typhoon Usagi in 09/21/2013 (orbit No. 90277)
at 02:09 UTC; and tropical storm Helene in 09/15/2006 (orbit No. 50338)
at 14:34 UTC. \label{fig:6}}
\end{figure}

\begin{figure}[H]
\noindent \begin{centering}
\includegraphics[width=0.74\paperwidth]{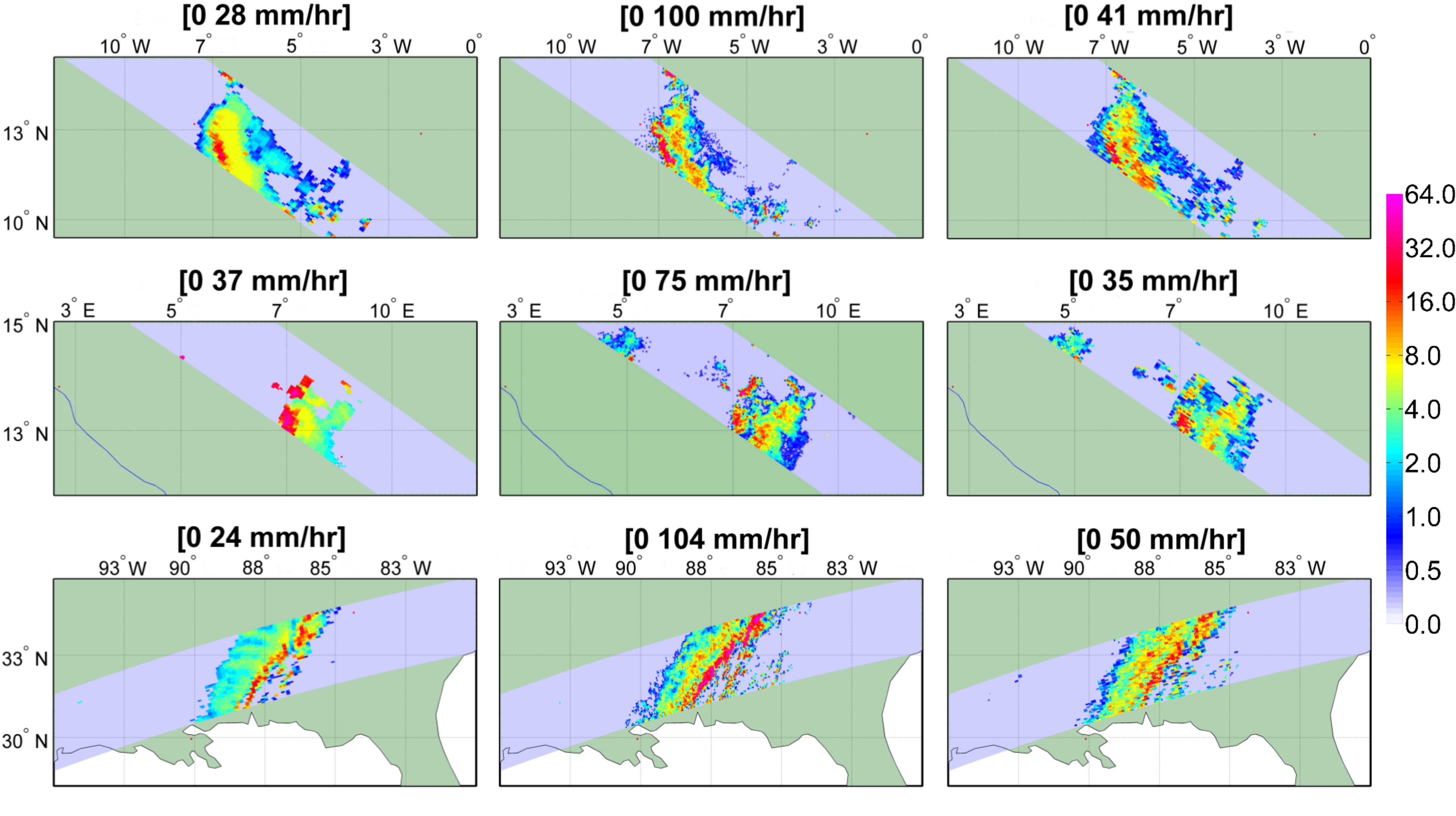}
\par\end{centering}
\protect\caption{From left to right: TMI-2A12, PR-2A25 and ShARP retrievals. Top-to-bottom
panels: a thunderstorm over Mali, Africa in 08/29/2010 (orbit No.
72841) at 10:30 UTC; a summertime thunderstorm over Nigeria, Africa
in 06/28/1998 (orbit No. 03357) at 17:43 UTC; and a spring season
squall line of precipitation supercells and tornadoes over Georgia,
U.S., in 01/30/2013 (orbit No. 86639) at 16:22 UTC.\label{fig:7}}
\end{figure}

\begin{figure}[h]
\noindent \begin{centering}
\includegraphics[width=0.74\paperwidth]{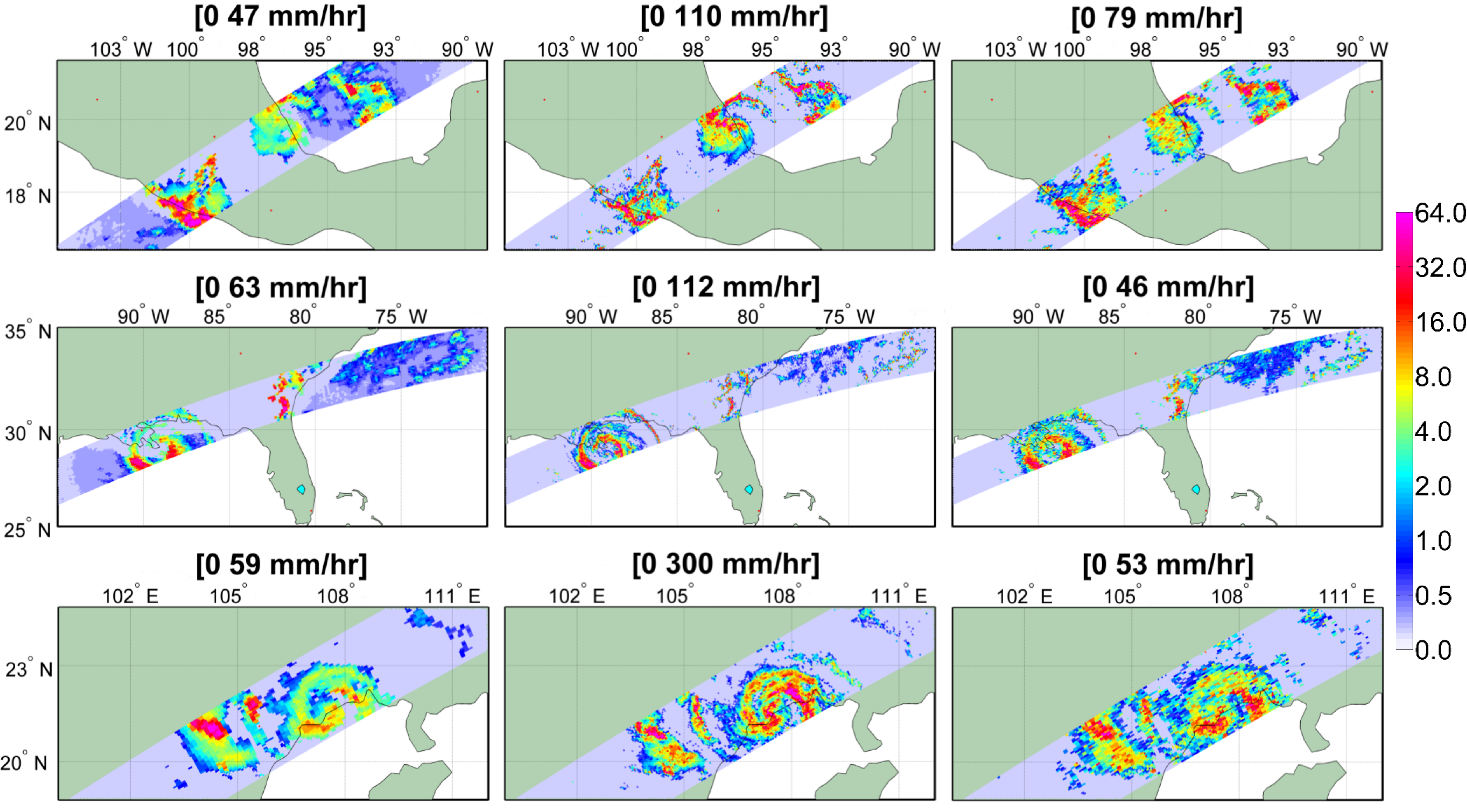}
\par\end{centering}

\protect\caption{From left to right: TMI-2A12, PR-2A25 and ShARP retrievals. Top to
bottom panels: tropical storm Fernand in 08/26/2013 (orbit No. 89874)
at 05:30 UTC, hurricane Isaac in 28/08/2012 (orbit No. 84227) at 22:12
UTC and typhoon Kai-tak in 08/17/2012 (orbit No. 84050) at 13:35 UTC.
\label{fig:8}}
\end{figure}

Figs.~\ref{fig:6}, \ref{fig:7} and \ref{fig:8} demonstrate the
results of few instantaneous retrieval experiments over ocean, land
and coastal areas, respectively. Here, we confined our consideration
to some important storms recorded in the TRMM extreme event archives
(\url{http://trmm.gsfc.nasa.gov/publications_dir/extreme_events.html}). 

Over ocean, we used the TMI snapshots of the hurricane Danielle (08/29/2010),
super typhoon Usagi (09/21/2013) and tropical storm Helene (09/15/2006)
(Fig.~\ref{fig:6}). Over land, we focused on a few thunderstorms
and mesoscale convective systems. These events include a squall line
over Mali (08/29/2010), a local thunderstorm over Nigeria (06/28/1998)
and a spring-season squall line containing tornadic activities over
Georgia, U.S. (01/30/2013) (Fig.~\ref{fig:7}). Over coastal areas,
we retrieved the TMI overpasses of the tropical storm Fernand over
the eastern coast of Mexico (08/26/2013), hurricane Issac over the
Mississippi delta U.S. (28, 29/08/2012) and typhoon Kai-tak over the
Gulf of Tonkin, coastlines of Vietnam and southern China (08/17/2012)
(Fig.~\ref{fig:8}).


\begin{figure}[H]
\noindent \begin{centering}
\includegraphics[width=0.65\paperwidth]{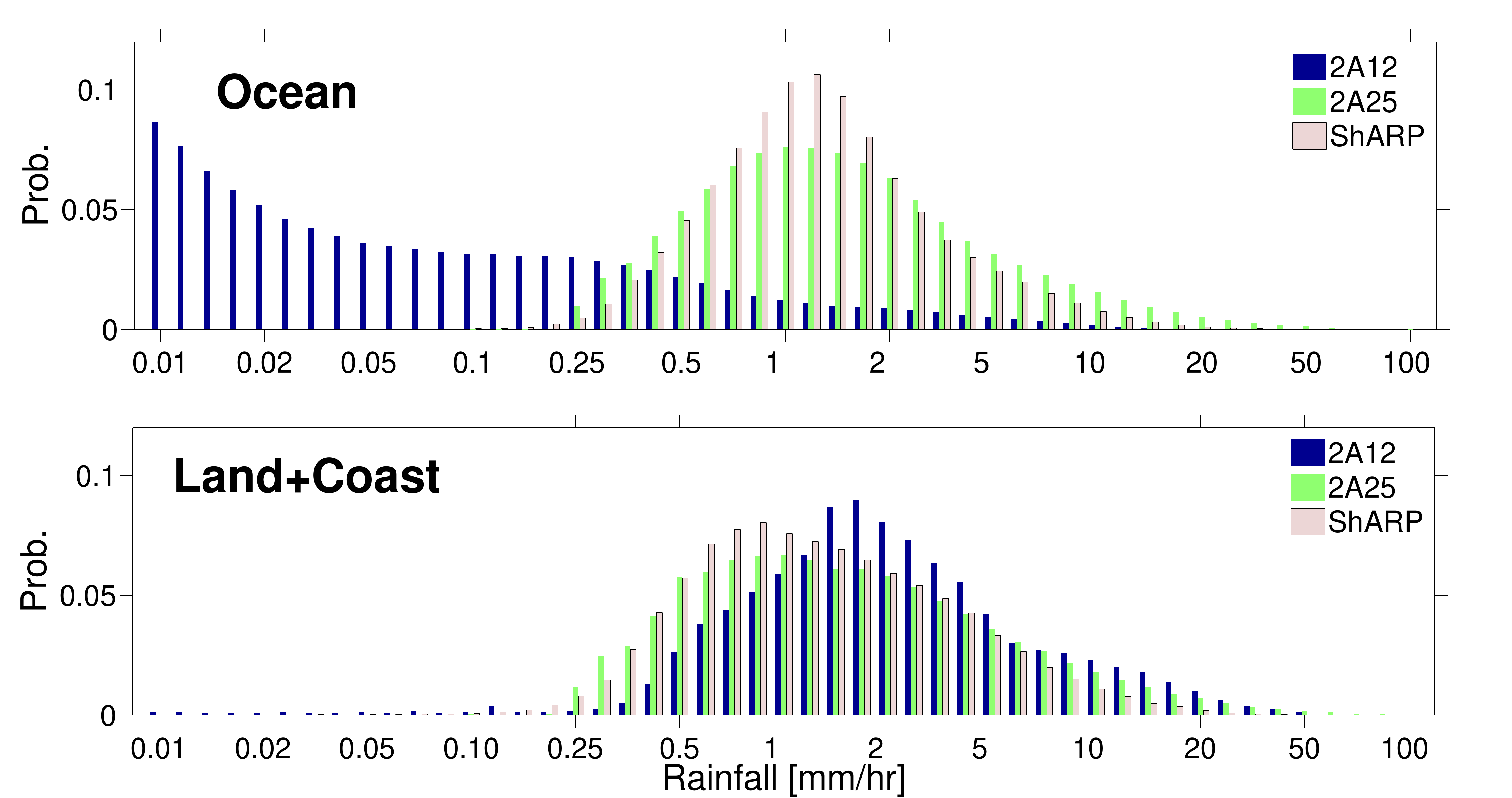}
\par\end{centering}

\protect\caption{Pixel level probability histograms of the instantaneous rainfall retrievals
(> 1e+6 points for each product) for the TMI-2A12 , PR-2A25 and ShARP
products over the ocean (top panel) and land-coast (bottom panel)
surface classes. \label{fig:9}}
\end{figure}

In general, our experiments in Fig.~\ref{fig:6}, \ref{fig:7}, and
\ref{fig:8} demonstrate good agreements between the ShARP retrieval
and the standard TRMM products. As previously noticed, we typically
see that the 2A12 retrieves much larger areas of light rain over ocean
compared to the 2A25 and thus to the current implementation of the
ShARP algorithm. We see that ShARP can properly recover the storm
morphology, high intense and light rainfall both over ocean and land.
For example, in the retrieval experiments of the tropical cyclones
over ocean (Fig.~\ref{fig:6}), the high-intensity rainfall cells,
curvature and multiband structure of the studied storms are well captured.
Over land, in the retrieved thunderstorm over Nigeria (Fig.~\ref{fig:6},
first row) and the frontal system over Georgia (Fig.~\ref{fig:6},
bottom row), we see that the ordinary cells and stratiform trailing
behind the leading edge of the squall line are well captured. Visual
inspections of the retrieved rainfall at the ocean-land interface
also confirm that the ShARP retrievals remain coherent over the interface
and are in good agreement with the 2A12 and 2A25. 

\begin{table}[h]
\noindent \begin{centering}
\begin{tabular}{|>{\centering}p{20mm}|c|c|c|c|c|c|}
\hline 
\multicolumn{7}{|c|}{\textbf{\small{}Retrieval Difference Metrics}}\tabularnewline
\hline 
\multirow{3}{20mm}{{\small{}Metrics}} & \multicolumn{6}{c|}{{\small{}Surface Classes}}\tabularnewline
\cline{2-7} 
 & \multicolumn{3}{c|}{{\small{}Ocean}} & \multicolumn{3}{c|}{{\small{}Land+Coast}}\tabularnewline
\cline{2-7} 
 & {\small{}(a)} & {\small{}(b)} & {\small{}(c)} & {\small{}(a)} & {\small{}(b)} & {\small{}(c)}\tabularnewline
\hline 
{\small{}RMSD} & {\small{}5.0} & {\small{}2.8} & {\small{}5.3} & {\small{}6.1} & {\small{}4.3} & {\small{}6.5}\tabularnewline
\hline 
{\small{}MAD} & {\small{}2.3} & {\small{}1.6} & {\small{}2.6} & {\small{}2.7} & {\small{}2.4} & {\small{}3.2}\tabularnewline
\hline 
{\small{}$\rho$} & {\small{}0.55} & {\small{}0.60} & {\small{}0.45} & {\small{}0.50} & {\small{}0.55} & {\small{}0.40}\tabularnewline
\hline 
\end{tabular}
\par\end{centering}

\protect\caption{Retrieval difference metrics obtained by comparing ShARP vs 2A25 (a),
ShARP vs 2A12 (b) and 2A12 vs 2A25 (c) for 100 randomly chosen orbital
tracks in 2013. Shown statistics are the Root Mean Squared Difference
(RMSD) {[}mm/hr{]}, Mean Absolute Difference (MAD) {[}mm/hr{]} and
Spearman's correlation ($\rho$). The statistics are obtained for
instantaneous rainfall estimates at pixel level over the intersection
of raining areas of all three retrieval products. \label{tab:3}}
\end{table}
Fig.~\ref{fig:9} compares the histogram of the retrieved rainfall
values at pixel-level, obtained from 100 randomly sampled orbits of
the TRMM in calendar year 2013. Overall, it is seen that the distribution
of ShARP and 2A25 are matched well. However, ShARP tends to retrieve
more rain around the mode and falls a bit short over the tail. This
behavior is somehow expected as ShARP uses a maximum a posteriori
estimator that implicitly seeks the mode of the rainfall distribution.
As previously noticed, over ocean, 2A12 retrieves much lower rain
rates than the other two products. In 2A12, the highest probable range
of rainfall intensity falls below 0.02 mm/hr. In effect, more than
75\% of raining cases are reported to be below 0.25 mm/hr, while the
probability of rainfall at this range is almost zero in the other
two products. In 2A25 and ShARP, 63 and 71\% of raining cases are
within the range of 0.25 to 0.5 mm/hr, respectively, while this probability
is around 0.2 in the 2A12. We see that the distribution of the 2A25
over ocean has the thickest tail among the others. In this product,
the probability of rainfall exceeding 10 mm/hr is \textasciitilde{}5\%,
while only 1.5\% and 0.7\% of raining cases are in this range for
ShARP and 2A12. Over land and coastal areas, the rainfall distributions
of all three products are more or less similar. The mode of the rainfall
is around 0.9 mm/hr in ShARP and 2A25, while the highest probable
rainfall values are concentrated around 1.9 mm/hr in the 2A12. It
is also apparent that the 2A25 and thus ShARP detect more lower rain
rates $<1$ mm/hr, while detection of higher rain rates $>10$ mm/hr
is more probable in the 2A12 over land. It is important to note that,
as the extent of raining areas are different in the studied retrievals,
the observed differences in the probability distribution of the instantaneous
rainfall do not necessarily lead to large differences in the volumetric
retrieved rainfall. In effect, we will show later on that the total
annual estimates of rainfall match well in all three products. 

\begin{figure}[h]
\noindent \begin{centering}
\includegraphics[width=0.6\paperwidth]{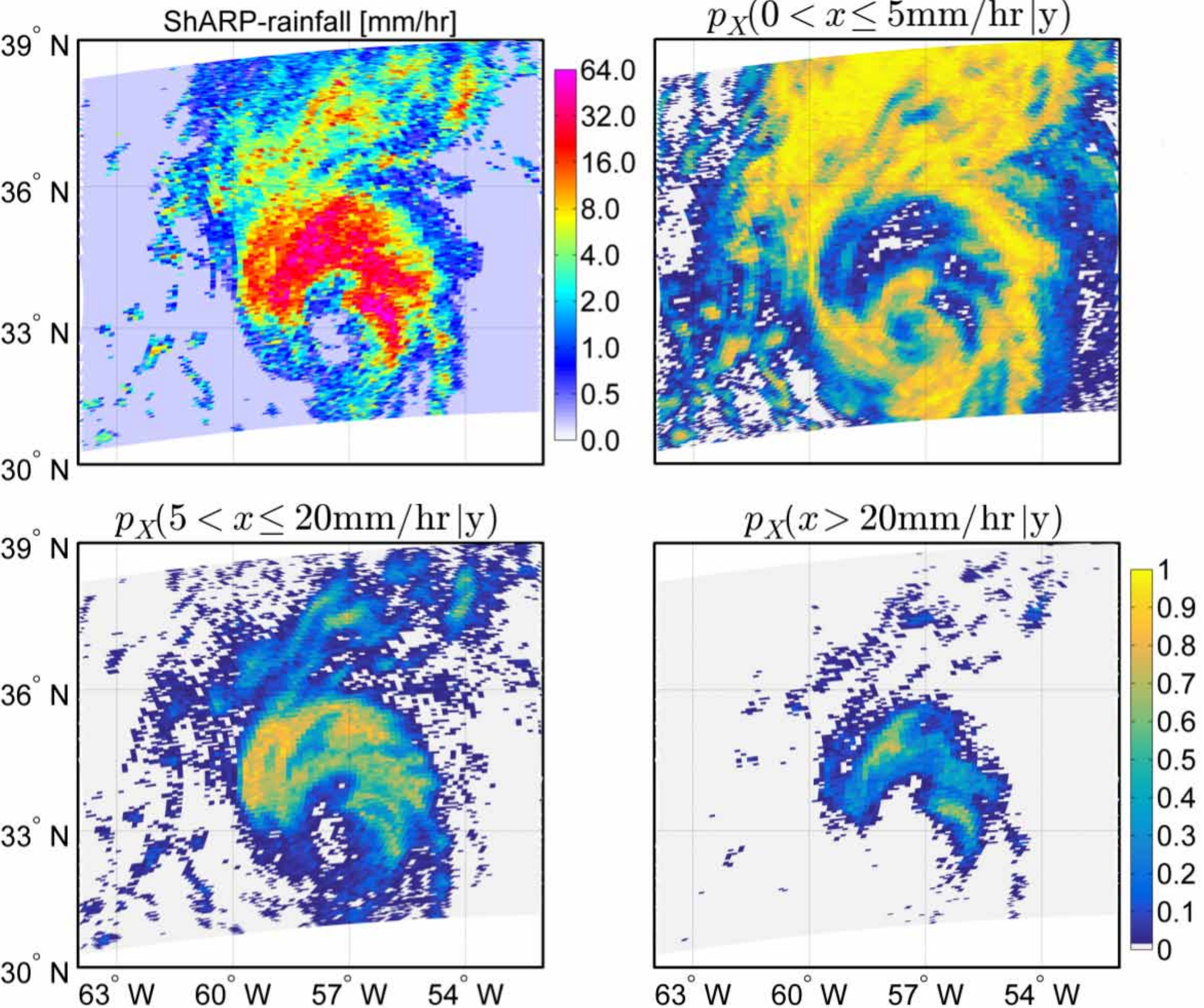}
\par\end{centering}

\protect\caption{Probability maps showing different segments of the posterior probability
density function $p_{X}\left(x|y\right)$ for the ShARP retrieval
(top left panel) of the hurricane Danielle (orbit No. 72840) at 09:48
UTC.\label{fig:10}}
\end{figure}

\begin{table}[H]
\noindent \begin{centering}
\begin{tabular}{|l|c|c|c|c|c|c|c|c|c|c|c|}
\hline 
\multicolumn{12}{|c|}{\textbf{\small{}Quantiles {[}mm/hr{]} of the ShARP Posterior PDF}}\tabularnewline
\hline 
\multirow{2}{*}{{\small{}Bins}} & \multirow{2}{*}{{\small{}Mean}} & \multicolumn{5}{c|}{{\small{}Ocean }} & \multicolumn{5}{c|}{{\small{}Land + Coast}}\tabularnewline
\cline{3-12} 
 &  & {\small{}5th} & {\small{}25th} & {\small{}50th} & {\small{}75th} & {\small{}95th} & {\small{}5th} & {\small{}25th} & {\small{}50th} & {\small{}75th} & {\small{}95th}\tabularnewline
\hline 
\hline 
{\small{}0.1-0.2} & {\small{}0.15} & {\small{}0.0} & {\small{}0.0} & {\small{}0.0} & {\small{}0.36} & {\small{}1.7} & {\small{}0.0} & {\small{}0.0} & {\small{}0.0} & {\small{}0.0} & {\small{}1.70}\tabularnewline
\hline 
{\small{}0.2-0.5} & {\small{}0.4} & {\small{}0.0} & {\small{}0.0} & {\small{}0.4} & {\small{}0.7} & {\small{}2.2} & {\small{}0.0} & {\small{}0.0} & {\small{}0.3} & {\small{}0.8} & {\small{}2.4}\tabularnewline
\hline 
{\small{}0.5-1.0} & {\small{}0.8} & {\small{}0.0} & {\small{}0.3} & {\small{}0.7} & {\small{}1.2} & {\small{}3.1} & {\small{}0.0} & {\small{}0.4} & {\small{}0.7} & {\small{}1.2} & {\small{}3.4}\tabularnewline
\hline 
{\small{}1.0-2.0} & {\small{}1.5} & {\small{}0.0} & {\small{}0.6} & {\small{}1.1} & {\small{}2.0} & {\small{}5.2} & {\small{}0.0} & {\small{}0.5} & {\small{}1.0} & {\small{}1.9} & {\small{}5.6}\tabularnewline
\hline 
{\small{}2.0-5.0} & {\small{}3.0} & {\small{}0.3} & {\small{}1.0} & {\small{}1.8} & {\small{}3.5} & {\small{}10.0} & {\small{}0.3} & {\small{}0.9} & {\small{}1.8} & {\small{}3.6} & {\small{}10.2}\tabularnewline
\hline 
{\small{}5.0-10.0} & {\small{}7.0} & {\small{}0.7} & {\small{}2.0} & {\small{}4.0} & {\small{}7.9} & {\small{}19.7} & {\small{}0.5} & {\small{}1.6} & {\small{}3.6} & {\small{}7.2} & {\small{}19.6}\tabularnewline
\hline 
{\small{}10.0-25.0} & {\small{}14.0} & {\small{}1.1} & {\small{}3.5} & {\small{}7.4} & {\small{}14.2} & {\small{}35.2} & {\small{}0.7} & {\small{}2.8} & {\small{}6.8} & {\small{}14.7} & {\small{}37.2}\tabularnewline
\hline 
{\small{}25.0-50.0} & {\small{}31.0} & {\small{}2.5} & {\small{}7.7} & {\small{}16.5} & {\small{}31.5} & {\small{}62.5} & {\small{}0.8} & {\small{}5.2} & {\small{}13.8} & {\small{}28.2} & {\small{}68.8}\tabularnewline
\hline 
\end{tabular}
\par\end{centering}

\protect\caption{Static estimation of key percentiles {[}mm/hr{]} of the posterior
PDF of the ShARP rainfall retrievals for 100 sampled orbits in calendar
year 2013. Second column denotes the mean {[}mm/hr{]} values of the
retrieved rainfall within each bin. \label{tab:4}}
\end{table}

To further validate the instantaneous retrieval of the ShARP algorithm,
we report the Root Mean Squared Difference (RMSD), Mean Absolute Difference
(MAD), and the Spearman's correlation ($\rho$) for each pair of the
studied products. Computation of these proximity measures for instantaneous
rainfall is not straightforward as these products do not share identical
sets of raining areas. Table~\ref{tab:3} shows pixel-level estimates
of these measures over the intersection of raining areas in the 100
randomly sampled orbits discussed in Fig.~\ref{fig:9}. As is evident,
ShARP and 2A12 are the closest pair while naturally ShARP is closer
and more correlated with the 2A25 than that of 2A12. Evaluating the
pixel level differences of the rainfall samples, among the studied
products, shows that typically a large number of those deviations
are very small while a small number of them are typically very large.
For instance, more than 55\% of the differences between ShARP and
2A25 are less than 1 mm/hr while less than 5\% of them are greater
than 8 mm/hr. This can be the main reason why the RMSD is almost twice
that of the MAD metric in Table~\ref{tab:3}. In effect, the RMSD
can be easily saturated by a few large deviations as it quadratically
penalizes them. On the other hand, MAD linearly penalizes the differences
and seems to be a more robust measure against a few number of large
deviations.

As we explained, the posterior density of the ShARP retrievals can
be empirically approximated via counting the frequency of rainfall
occurrence in the atoms of the rainfall sub-dictionaries. Table~\ref{tab:4}
reports a static estimation of key percentiles of the posterior PDF
for the examined 100 orbits. For brevity, we only present the results
for the rainfall values falling between 0.1, 0.2, 0.5, 1, 2, 5, 10,
25 and 50 mm/hr. Fig.~\ref{fig:10} also shows some dynamic probability
maps of the posteriori PDF for the snapshot of the hurricane Danielle
shown in the Fig.~\ref{fig:6}. Clearly, this important feature of
ShARP allows us to perform rainfall retrieval probabilistically and
track the high risk areas of the extreme rainfall based on a certain
probability of exceedance. 

\begin{figure}[h]
\noindent \begin{centering}
\includegraphics[width=0.75\paperwidth]{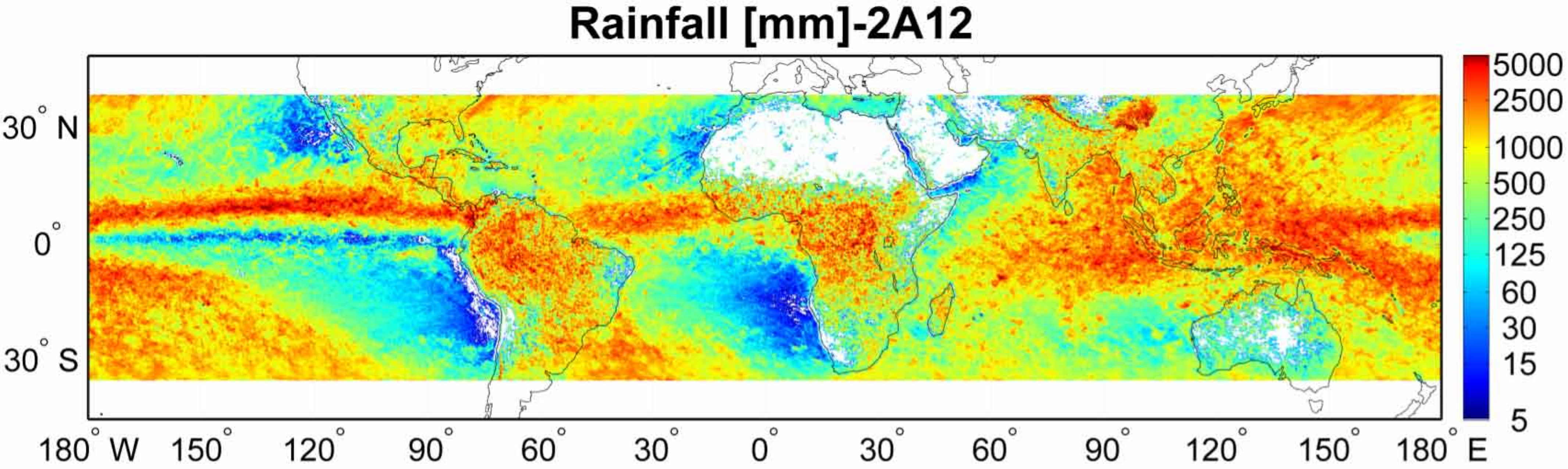}
\par\end{centering}

\noindent \begin{centering}
\includegraphics[width=0.75\paperwidth]{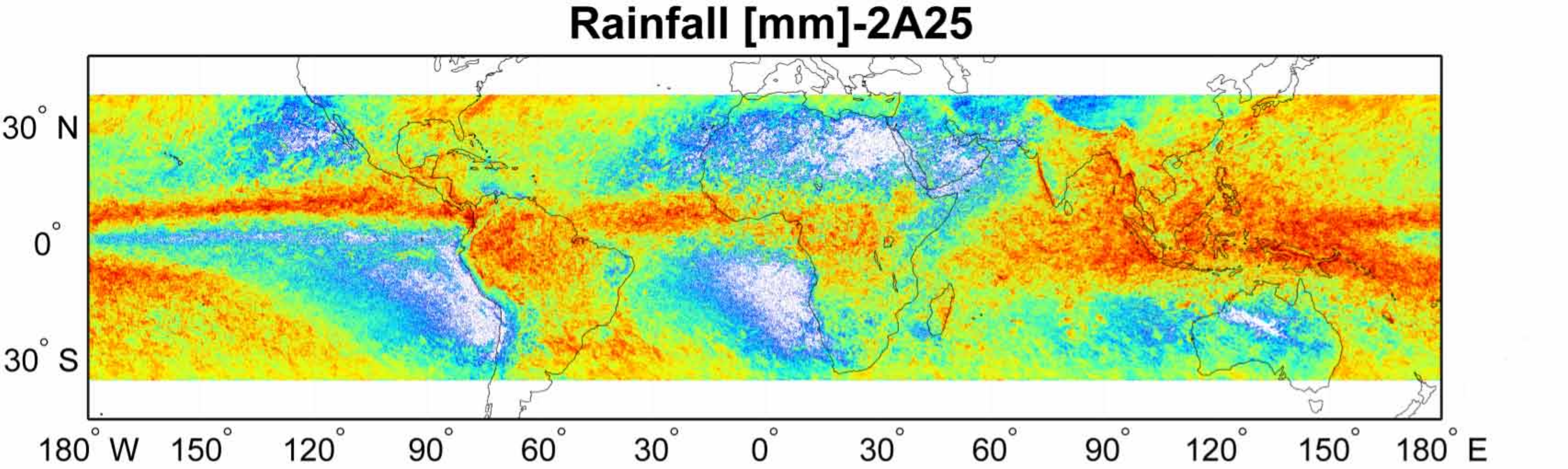}
\par\end{centering}

\noindent \begin{centering}
\includegraphics[width=0.75\paperwidth]{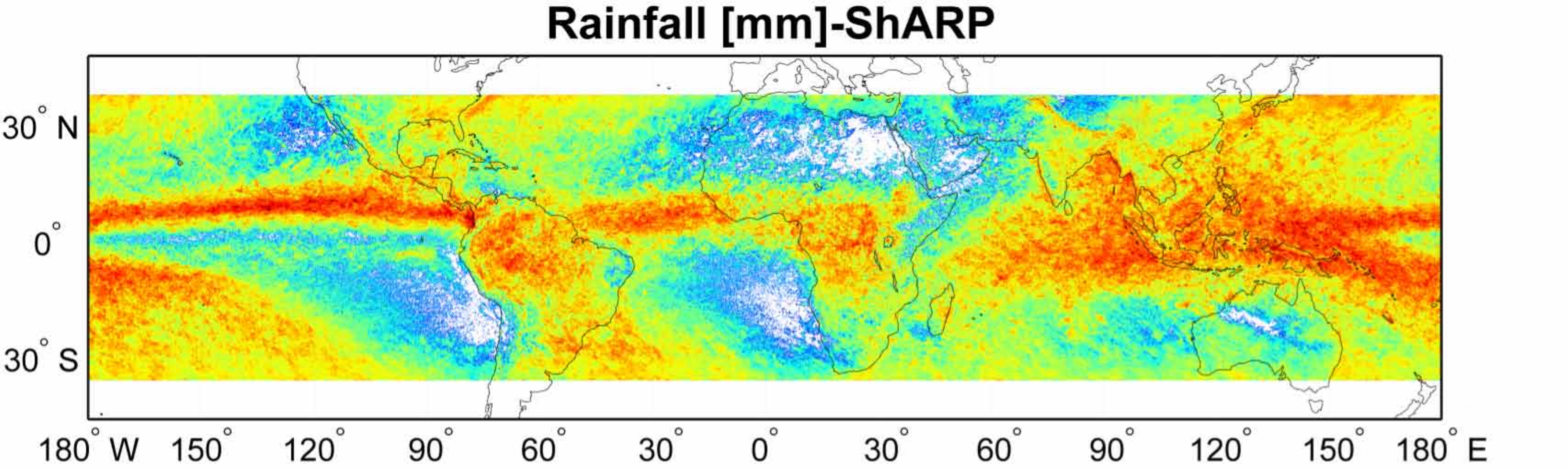}
\par\end{centering}

\protect\caption{Annual estimates of the total rainfall {[}mm{]} in 2013 mapped onto
a 0.1-degree grid box. From top to bottom panels: 2A12, 2A25, and
ShARP retrieval products. \label{fig:11}}
\end{figure}

\subsection{Cumulative Experiments}

To validate the results of our algorithm in a cumulative sense, we
focus on all orbital observations of the TRMM in calendar year 2013.
To unify the sampling rate, we only use the available observations
over the inner swath, where both sensors provide overlapping and validated
rainfall observations. Fig.~\ref{fig:11} demonstrates the annual
rainfall estimates, mapped onto a $0.1^{\circ}\times0.1^{\circ}$
grid. In general, we see a good agreement between ShARP and the standard
TRMM products. Here, as the 2A25 potentially provides one of the best
spaceborne estimates of the total rainfall volume over the tropics
\citep[see,][]{Berg2009}, we also study deviations of the passive
retrievals from this active product.

\begin{figure}[h]
\noindent \begin{centering}
\includegraphics[width=0.75\paperwidth]{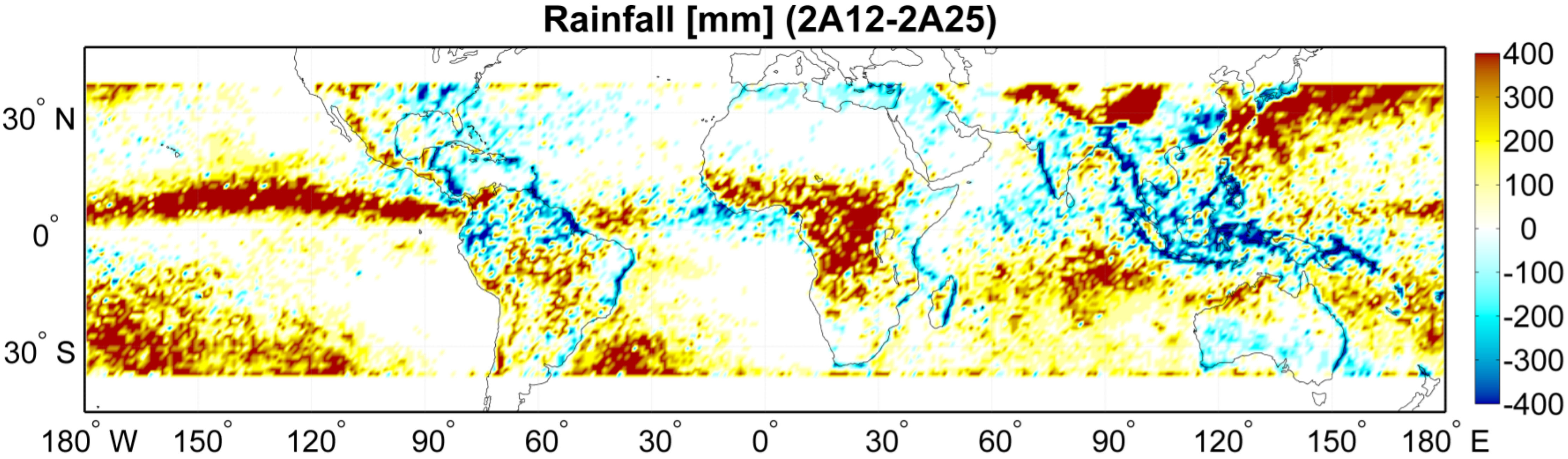}
\par\end{centering}

\noindent \begin{centering}
\includegraphics[width=0.75\paperwidth]{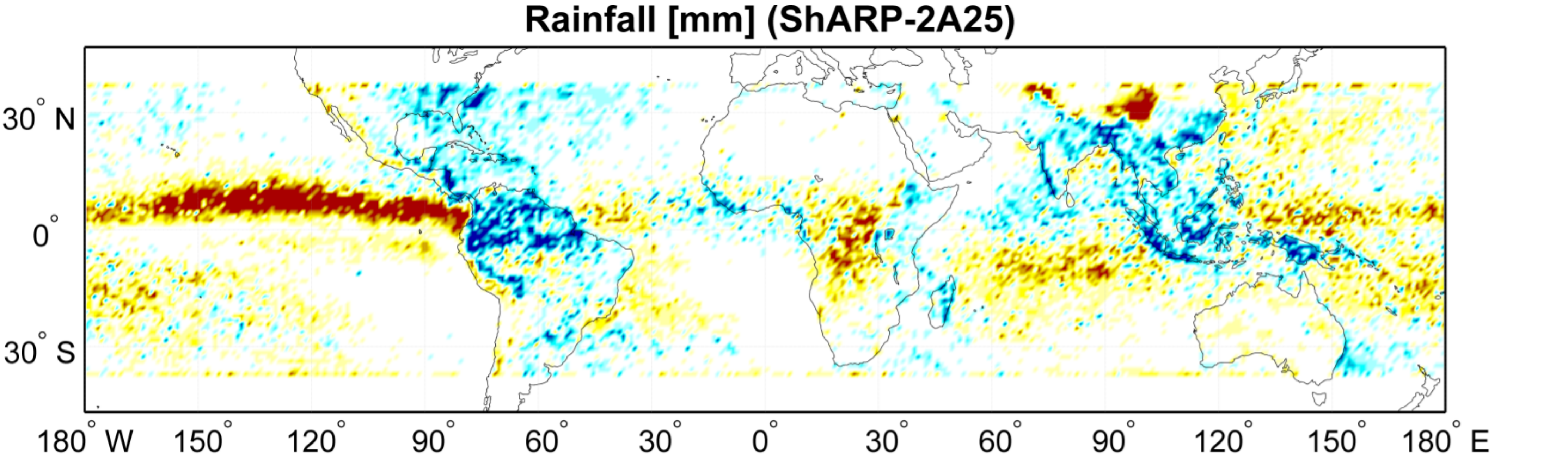}
\par\end{centering}

\protect\caption{Annual estimates of the total rainfall difference {[}mm{]} for calendar
year 2013. From top to bottom panels: The difference between the 2A12,
and ShARP retrievals with the 2A25 at grid size $1^{\circ}\times1^{\circ}$.
Hot (red) and cold (blue) colors denote intensity of positive and
negative differences. \label{fig:12}}
\end{figure}

\begin{figure}[H]
\noindent \begin{centering}
\includegraphics[width=0.75\paperwidth]{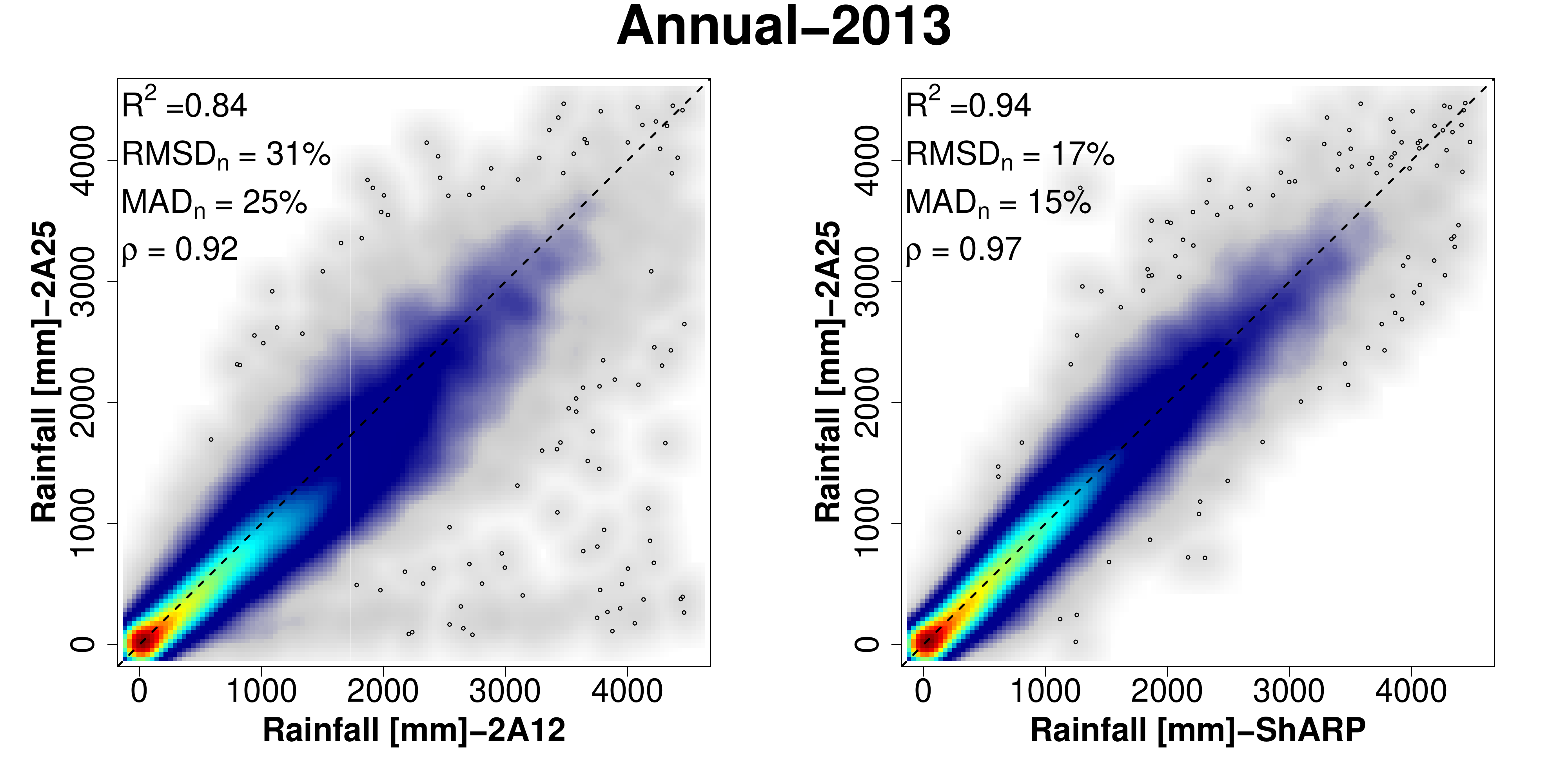}
\par\end{centering}

\protect\caption{Smooth scatter plots of the annual retrieved rainfall {[}mm{]} by
the 2A12 (left panel) and ShARP (right panel) versus the 2A25 at grid
size $1^{\circ}\times1^{\circ}.$ Hot (red) and cold (blue) colors
denote higher and lower density of the available rainfall intensity
pairs. The R-square denotes the coefficient of determination, $\text{RMSD}_{\text{n}}$
and $\text{MAD}_{\text{n}}$ are the normalized root mean squared
and mean absolute difference while $\rho$ denotes the correlation
coefficient. \label{fig:13}}
\end{figure}

\begin{figure}[H]
\noindent \begin{centering}
\includegraphics[width=0.7\paperwidth]{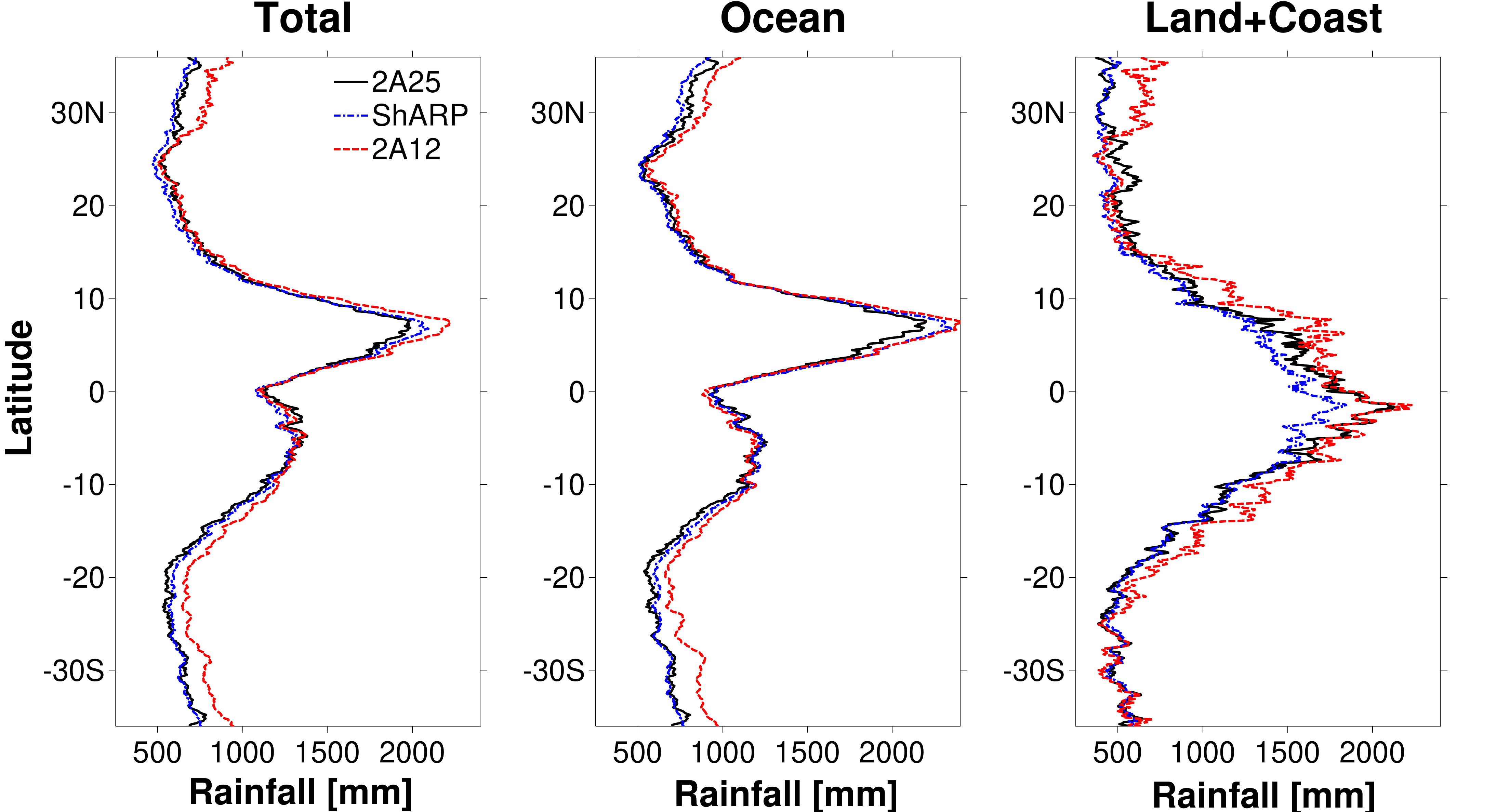}
\par\end{centering}

\protect\caption{Annual rainfall zonal mean values {[}mm{]} obtained from estimates
of the annual rainfall shown in Fig.~\ref{fig:11}. From left to
right panels: zonal mean values computed over all surface classes,
over ocean and over land-coasts. \label{fig:14}}
\end{figure}

At $0.1$-degree resolution, the normalized root mean squared difference
($\text{RMSD}_{\text{n}}$)%
\footnote{$\text{RMSD}_{\text{n}}$ is the RMSD, which is normalized by the
square root of the sum of squared of the reference field at a pixel
level.%
} is about 36\% and 48\% for ShARP (Fig.~\ref{fig:11}, bottom panel)
and 2A12 (Fig.~\ref{fig:11}, top panel), respectively. At the coarser
resolution of $1^{\circ}\times1^{\circ}$ grid box, this metric reduces
to 17\% and 31\% (Fig.~\ref{fig:12}), while the overall correlation
with 2A25 is 0.92 and 0.97 for the 2A12 and ShARP (Fig.~\ref{fig:13}).
Zonal mean values are also demonstrated in Fig.~\ref{fig:14} with
quantitative explanations in Table~\ref{tab:5}. Over ocean, except
in the North Atlantic mid-latitude storm tracks, both ShARP and 2A12
slightly overestimate the total rainfall obtained from 2A25, while
most of the underestimation regions are spread over land, especially
near coastal zones, islands and peninsulas, although some overestimation
can be seen in the Central Africa and South America in both 2A12 and
ShARP. 

Fig.~\ref{fig:12} shows that passive retrieval products overestimate
(\textasciitilde{} 300-400 mm) 2A25 on the narrow ridge of high precipitation
in the Intertropical Convergence Zone (ITCZ) across the Pacific Ocean.
As is evident, over the South Pacific, Atlantic, and Indian Ocean
convergence zones, we also see some over estimation in ShARP, while
the positive difference is relatively mitigated compared to the standard
2A12. In the North Atlantic mid-latitude storm tracks, both passive
retrievals slightly underestimate the annual rainfall while the deviations
are smaller in the 2A12 compared to ShARP.

Some promising results of our algorithm seem to be over land and coastal
zones. Over the subtropical hot desert, arid and semi-arid climates
(e.g., Sahara, Arabian, Syrian deserts, and central Iran plateau),
we see that ShARP retrieves well the low rainfall amounts seen by
the PR. Over the Central Africa, both 2A12 and ShARP overestimate
the 2A25 annual rainfall while the gap seems to be smaller in ShARP.
Over the North America, it is seen that 2A12 shows good agreement
with the PR estimates over the East Coast and Midwest of the United
States. However, ShARP approximates well the PR over the West Coast
and Southwest where the rainfall signatures are predominantly corrupted
with noise due to the highly emissive desert surfaces. Over South
America, ShARP shows improved retrieval over Brazil and southern Amazon,
while, compared to the 2A12, notable underestimation can be seen over
the northern Amazon basin, Colombia and Venezuela. Some improved results
of our algorithm are over the snow-covered Tibetan highlands and Himalayas.
We can see that ShARP can distinguish well the background noise from
rainfall signatures and reduces some overestimation seen in the 2A12.
Note that, we have used minimal number of the earth surface classifications
and have not defined any specific class over the Tibetan Plateau.
Indeed, due to the 9-dimensional nearest neighborhood selection of
the spectral sub-dictionaries, our algorithm is apparently capable
to robustly eliminate a large portion of the physically inconsistent
spectral candidates in the detection step. Over Southeast Asia, where
the rainfall signatures are masked by a mixture of ocean and land
surface background radiation regimes, both ShARP and 2A12 underestimate
the 2A25. However, the negative differences in ShARP are slightly
reduced, compared to the 2A12, especially over Indonesia, Malaysia
and Philippines. 

Comparison of the total annual zonal mean values (Fig.~\ref{fig:14},
left panel) shows that ShARP approximates well the average latitudinal
rainfall distribution and budget. We can see that not only over the
tropics but also over the mid-latitudes, where stratiform rainfall
is dominant, ShARP well reconstructs the 2A25 product over ocean (Fig.~\ref{fig:14},
middle panel). Over land, ShARP underestimates the zonal mean within
a narrow band (latitudes $5{}^{\circ}$S-N) around the tropics, while
it performs well over the subtropical climate zones (Fig.~\ref{fig:14},
right panel). This underestimation is contributed mainly by the ShARP
poor retrieval skill over the northern part of the South America.
Quantitative comparison of these zonal profiles is presented in Table~\ref{tab:5}.

\begin{table}[t]
\noindent \begin{centering}
{\small{}}%
\begin{tabular}{|c|l|>{\centering}m{20mm}|>{\centering}m{20mm}|}
\hline 
\multirow{2}{*}{\textbf{\small{}Product}} & \multirow{2}{*}{\textbf{\small{}Surface Class}} & \multicolumn{2}{c|}{\textbf{\small{}Annual Zonal Mean }}\tabularnewline
\cline{3-4} 
 &  & {\small{}RMSD} & MD\tabularnewline
\hline 
\hline 
\multirow{3}{*}{\textbf{\small{}ShARP-2A25}} & {\small{}Total } & {\small{}40.20} & {\small{}-6.53}\tabularnewline
\cline{2-4} 
 & {\small{}Ocean} & {\small{}47.61} & {\small{}7.15}\tabularnewline
\cline{2-4} 
 & {\small{}Land + Coast} & {\small{}95.67} & {\small{}-41.02}\tabularnewline
\hline 
\multirow{3}{*}{\textbf{\small{}2A12-2A25}} & {\small{}Total } & {\small{}103.04} & {\small{}73.63}\tabularnewline
\cline{2-4} 
 & {\small{}Ocean} & {\small{}99.50} & {\small{}69.42}\tabularnewline
\cline{2-4} 
 & {\small{}Land + Coast} & {\small{}137.62} & {\small{}79.43}\tabularnewline
\hline 
\end{tabular}
\par\end{centering}{\small \par}

\protect\caption{Retrieval skills including RMSD {[}mm{]} and mean difference (MD)
{[}mm{]} for the annual zonal mean values shown in Fig.~\ref{fig:14}.
\label{tab:5} }
\end{table}

\begin{figure}[H]
\noindent \begin{centering}
\includegraphics[width=0.6\paperwidth]{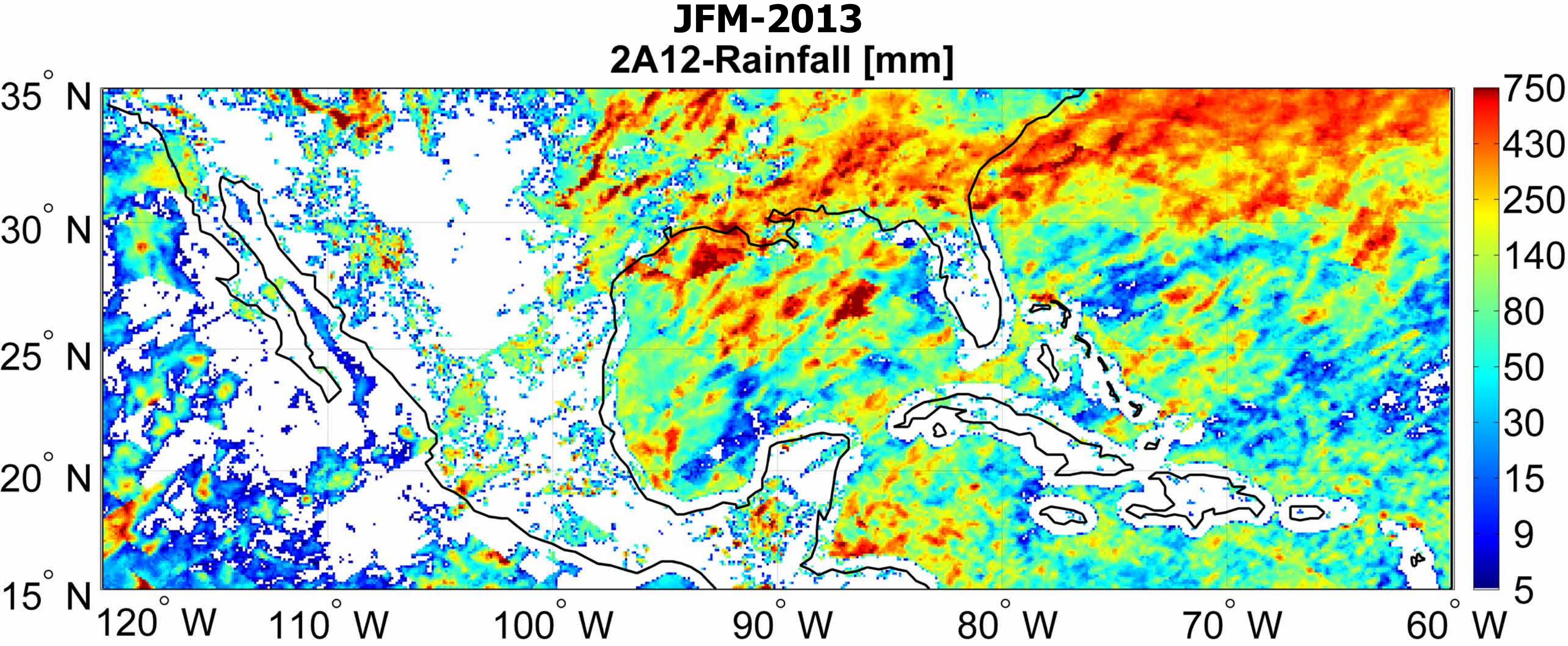}
\par\end{centering}

\noindent \begin{centering}
\includegraphics[width=0.6\paperwidth]{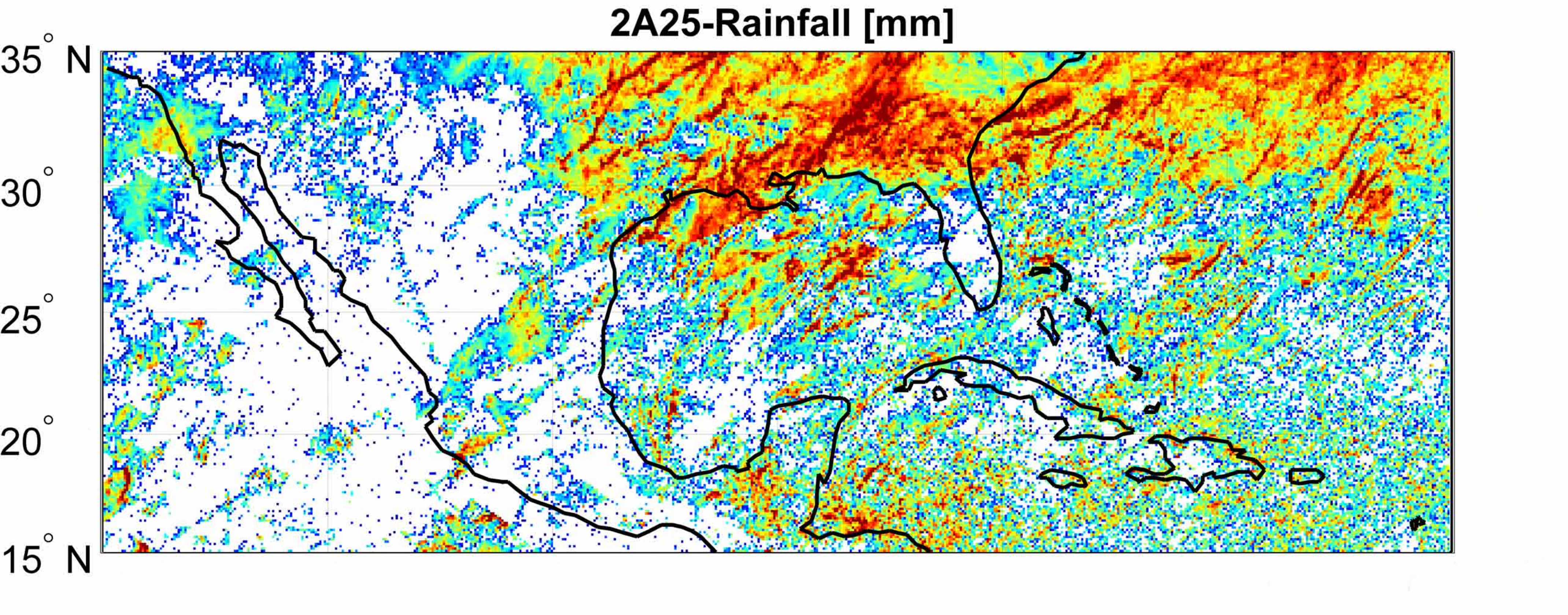}
\par\end{centering}

\noindent \begin{centering}
\includegraphics[width=0.6\paperwidth]{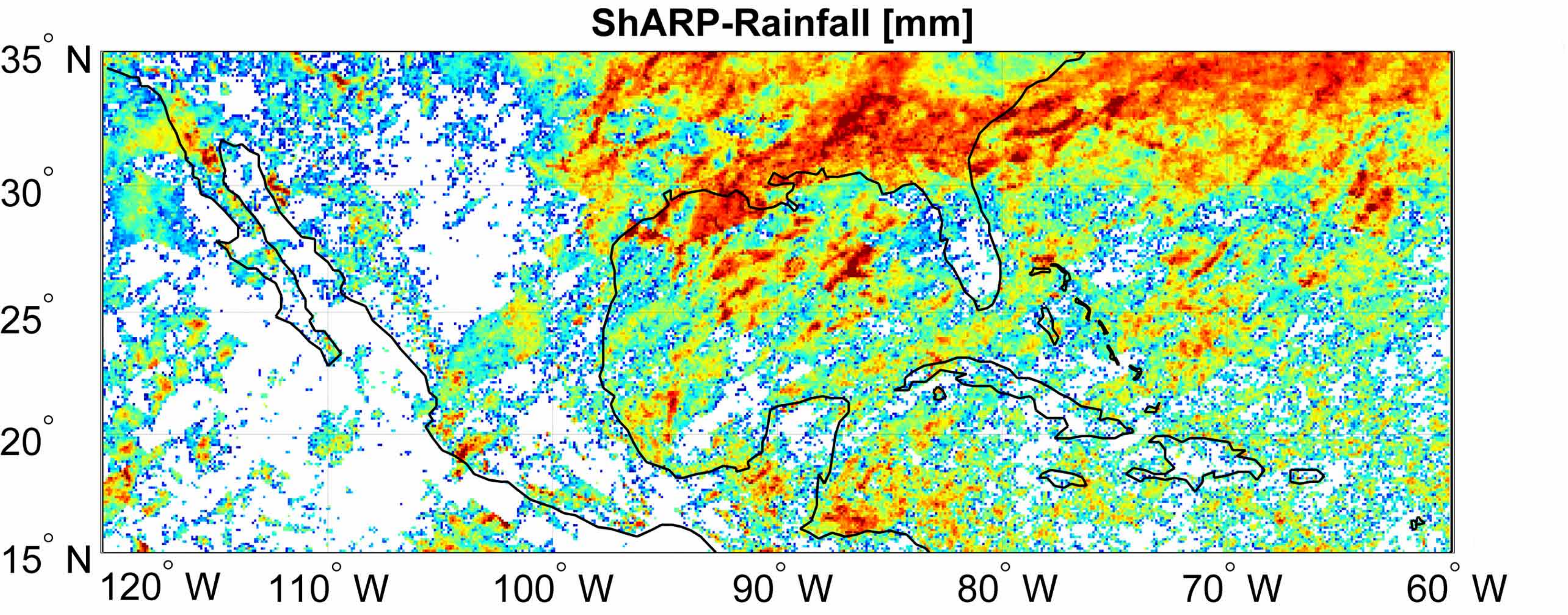}
\par\end{centering}

\protect\caption{Inter-annual rainfall accumulation for the period from January through
March (JFM) in 2013, mapped onto a 0.1-degree grid box.\label{fig:15}}
\end{figure}

\begin{figure}[h]
\noindent \begin{centering}
\includegraphics[width=0.75\paperwidth]{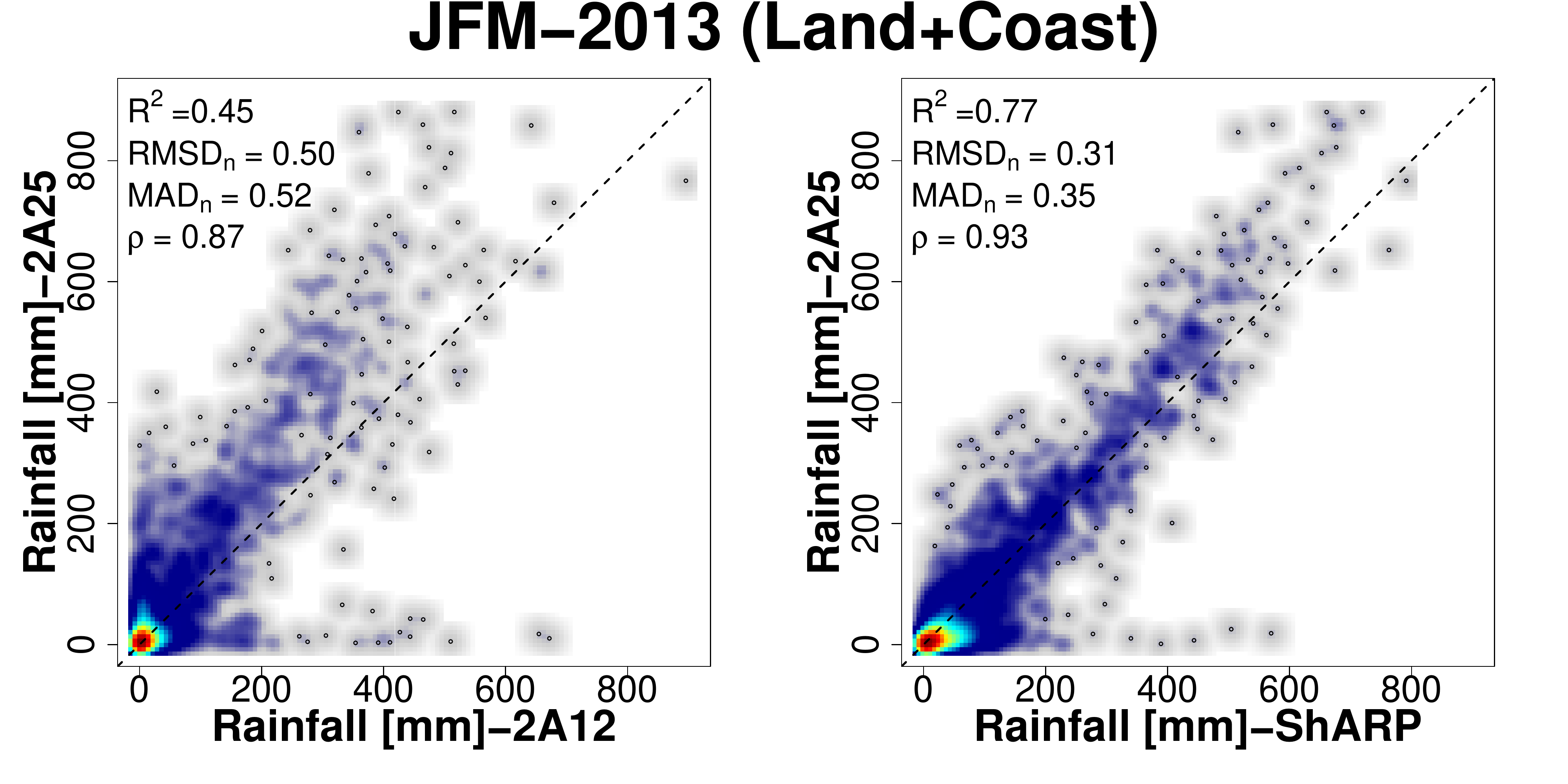}
\par\end{centering}

\protect\caption{Smooth scatter plots of the 3-month rainfall accumulation {[}mm{]},
shown in Fig.~\ref{fig:15}. The plots show the 2A12 (left panel)
and ShARP (right panel) versus the 2A25 at $0.5\times0.5^{\circ}$
grid box. See caption of Fig.~\ref{fig:13} for explanations of the
presented statistics. \label{fig:16}}
\end{figure}

To briefly evaluate inter-annual performance of our algorithm, especially
over land and coastal areas, we also focused on a 3-month rainfall
accumulation for the period from January through March (JFM) of 2013.
We confined the spatial extent of our evaluation within latitudes
$15$-$35{}^{\circ}$N and longitudes $60$-$120{}^{\circ}$W (Fig.~\ref{fig:15}).
The storm system of the area in the JFM period is mainly supplied
by the moisture coming from the Pacific Ocean through the subtropical
jet stream and is intensified where the extratropical lifting saturates
the atmospheric column over the Gulf of Mexico. This mechanism typically
causes heavy precipitation events over the southeast of the United
States and the Gulf of Mexico, while it leaves the southwest relatively
dry. Overall, we see that ShARP properly retrieves the high and low
seasonal precipitation amounts in the JFM system and its retrieved
rainfall resembles well the standard TRMM products. Specifically,
it is seen that in the vicinity of coast lines of the Caribbean Islands
and Bahamas, the light rainfall values are well captured by ShARP.
During this period, consistent with the instantaneous results shown
in Fig.~\ref{fig:5}, the largest amount of raining areas over the
ocean is detected by the 2A12 (88\% of ocean), while this fraction
is 71 and 66\% in the ShARP and 2A25. In contrast, ShARP detects the
largest raining area (69\%) over land, while this fraction is 63 and
50\% in the 2A25 and 2A12 . The main factor contributing to the overestimation
of the 2A25 by ShARP is primarily due to the coarse resolution of
the TMI sensor which is unable to resolve signatures of small-scale
precipitation events captured by the PR. A brief quantitative comparison
of the JFM rainfall system, only over land and coastal areas, is presented
in Fig.~\ref{fig:16}. As is evident, ShARP well correlates with
the 2A25, while we see some discrepancies showing that for some light
raining areas in the 2A25, both ShAPR and 2A12 retrieve high rainfall
values. It turns out that some of these anomalies are due, in part,
to miss interpretation of the highly emissive ground as rainfall signatures.
For example, we see that over the Baja California Desert, ShARP exhibits
over estimation spots while snow-covered land surface in the month
of January confuses the 2A12 algorithm over the north west of Arizona
(\textasciitilde{}$110{}^{\circ}$W, $35^{\circ}$N).

\section{CONCLUDING REMARKS\label{sec:5}}

We proposed a Bayesian microwave rainfall retrieval algorithm that
makes use of a priori collected rainfall and spectral dictionaries.
This algorithm relies on a nearest neighborhood detection rule and
exploits modern shrinkage estimation paradigms. We have examined its
performance using empirical dictionaries populated from coincidental
observations of the TRMM precipitation radar (PR) and Microwave Imager
(TMI) and demonstrated its considerable promise to provide accurate
rainfall retrievals especially over land and coastal areas. In future
research, the algorithm needs to be further verified for different
rainfall regimes over ocean and land. Further efforts also need to
be devoted for improving the retrieval of rainfall extremes both over
land and ocean. While we have confined our experiments to empirical
dictionaries, the core of our algorithm is flexible and versatile
enough to exploit both observational and physically-based generated
dictionaries. The proposed implementation is very parsimonious at
this stage and further refinements, such as smarter choices of surface
classes by considering ground emissivity patterns and adding auxiliary
state variables to the dictionaries (e.g., surface skin temperature,
total column water) can definitely improve performance of the proposed
approach. Currently, we are developing a new version of this algorithm
that uses compact dictionaries for faster and more accurate retrieval
of the entire rainfall profile. The particular emphasis will be on
the available spectral bands (10.65 to 183 GHz) of the radiometer
and observations of the dual frequency precipitation radar aboard
the successfully launched Global Precipitation Measuring (GPM) satellites.

\section*{ACKNOWLEDGMENT}

First author would like to thank Professor Christian D. Kummerow for
his advice at the early stage of this research. This work was supported mainly
by a NASA Earth and Space Science Fellowship under the contract NNX12AN45H,
the K. Harrison Brown Family Chair, and the Ling Endowed Chair funding.
Furthermore, the support provided by two NASA Global Precipitation
Measurement grants and the Belmont Forum DELTAS grant under the contract
NNX13AG33G, NNX13AH35G and ICER-1342944 are also greatly recognized. The TRMM 2A12 and 2A25 data were obtained
through the anonymous File Transfer Protocol publicly available at
\url{ftp://trmmopen.gsfc.nasa.gov/pub/trmmdata}.

\bibliographystyle{IEEEtranS}

\end{document}